\documentclass[12pt]{article}
\usepackage[margin=0.5in]{geometry}
\usepackage{amsfonts}
\usepackage{amsmath}
\usepackage{lscape}
\usepackage{graphicx}
\usepackage{epstopdf}
\usepackage{color}
\usepackage{float} 
\usepackage{arydshln} 
\usepackage[lined,boxed,commentsnumbered]{algorithm2e} 
\usepackage{caption} 
\usepackage{subcaption} 
\usepackage{amsthm} 
\usepackage{multirow} 

\usepackage{secdot}
\usepackage{amssymb}
\usepackage{setspace} 

\begin{document}

\title{Network Inference from Grouped Data}
\date{}
\author{
    YUNPENG ZHAO\\
    George Mason University\\
    yzhao15@gmu.edu \and
    CHARLES W. WEKO\\
    United States Army \\
    charles.w.weko.mil@mail.mil
}

\maketitle

\begin{abstract}
	
In medical research, economics, and the social sciences data frequently appear as subsets of a set of objects. Over the past century a number of descriptive statistics have been developed to construct network structure from such data.  However, these measures lack a generating mechanism that links the inferred network structure to the observed groups. To address this issue, we propose a model-based approach called the \textit{Hub Model} which assumes that every observed group has a leader and that the leader has brought together the other members of the group. The performance of Hub Models is demonstrated by simulation studies. We apply this model to infer the relationships among Senators serving in the 110$^{th}$ United States Congress, the characters in a famous 18$^{th}$ century Chinese novel, and the distribution of flora in North America.
	
\end{abstract}

{\bf Keywords:} Social network analysis, affiliation network, expectation-maximization algorithm, half weight index, Dream of the Red Chamber.

	\section{INTRODUCTION}\label{S:Introduction}
	
	A network of $n$ objects can be denoted by $N=(V,E)$, where $V=\{v_1,v_2,...,v_n \}$ is the set of nodes, and $E$ is the set of edges between nodes. In this article, we focus on symmetric weighted networks represented by an $n \times n$ adjacency matrix, $A$, where the element $A_{ij}$ measures the strength of the relationship between nodes $v_i$ and $v_j$. 
	
	Traditionally, statistical network analysis focuses on modeling \textit{observed} network structure (e.g.,  highway systems or electrical transmission grids).  In this situation, nodes are well defined and the physical links between these nodes can be directly observed \cite{Hiller01,Newman11}. However, in some fields of research the explicit network structure may not be observable.  This is particularly true of the social sciences.  In these fields, the observable data are groups of individuals and it is presumed that the groups are the result of some latent network. The fundamental task is to construct a network from such data.
	
	
	A familiar example is provided by Wasserman and Faust \cite{Wasserman94} who introduce the problem of inferring the relationships among a collection of children based on their attendance at birthday parties. In this introductory dataset, the children represent nodes in the network and the birthday parties represent subsets of children.
	
	
	In our paper, a collection of nodes observed in the same sample is called a \textit{group} and a dataset of these observations is referred to as \textit{grouped data}.  In Wasserman and Faust's example, each party defines a group and the set of all parties is the grouped data. Two individuals are said to \textit{co-occur} if they appear in the same group.

	One common technique used to estimate $A$ from grouped data is to count the number of times that a pair of nodes appears in the same group \cite{Zachary77,Freeman89,Wasserman94,Kolaczyk09,Brent11}.  Frequently, a threshold is applied to this count to create a symmetric unweighted undirected adjacency matrix; however, it has been shown that the characteristics of networks inferred by this technique are sensitive to the choice of threshold \cite{Choudhury10}.  Therefore, we adopt a generalized version of the  inter-citation frequency \cite{Kolaczyk09} which measures the number of times a pair of nodes is observed to co-occur in the dataset.  We refer to this measure as the \textit{co-occurrence matrix}. 
	
	
	
	An alternative technique, called the \textit{half weight index} \cite{Cairns87,Bejder98}, estimates $A$ by the frequency that two nodes co-occur given that one of them is observed.  This addresses a shortcoming of the co-occurrence matrix in which nodes that appear rarely can be estimated to have a weak relationship even though the relationship is quite strong \cite{Cairns87,Voelkl11}.
	
	
	
	The co-occurrence matrix and half weight index both have probabilistic interpretations. The co-occurrence matrix estimates the probability that two nodes will be observed together.  The half weight index estimates the probability that two nodes will be observed together given that one of them is observed. However, the probability or conditional probability of co-occurrences is not equivalent to the probability of connection of nodes. In fact, neither of these techniques describe the process which leads to the generation of the observed groups. It is unclear how the descriptive statistics relate to the grouped data in these methods.
	
\begin{figure}[H]
		\begin{center}
			\fbox{
				\includegraphics{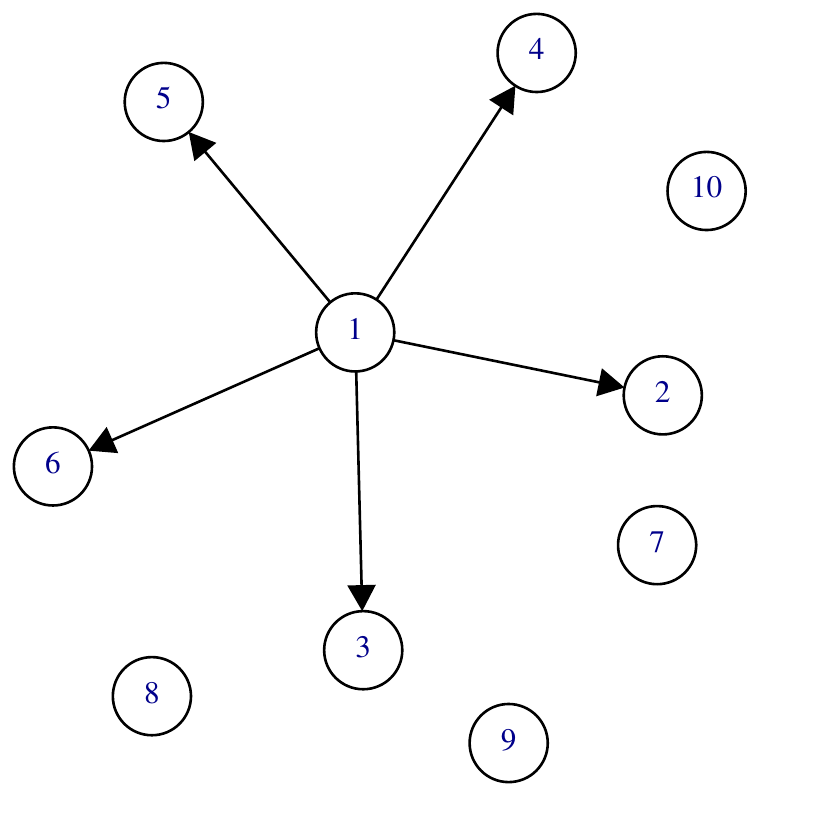}
			}
		\end{center}
		\caption{The generating mechanism of the Hub Model is demonstrated on a group of 10 nodes. In the observed sample, nodes $v_1,\dots,v_6$ are members of the group while nodes $v_7,\dots,v_{10}$ are not members of the group.  The observed group is the result of the hub node, $v_1$, bringing together nodes $v_2,\dots,v_6$.}
		\label{F:Sample Hub}
	\end{figure}

	We propose a model-based approach to infer the latent network from grouped data. We refer to this model of group formation as the \textit{Hub Model} because each observed group is assumed to be brought together by a hub node.  A group generated by the Hub Model is illustrated in Figure \ref{F:Sample Hub}.

	The Hub Model belongs to the family of finite mixture models. Such mixture models  and their variants have been applied in many different situations including text classification  \cite{Carreira00}, topic models \cite{Anandkumar15}, fingerprint identification \cite{Vretos12}, and product recommendation \cite{Colace15}.
	
	One of the advantages of Hub Models is that we are able to give the strength of the relationship $A_{ij}$ a definition which is both mathematically clear and at the same time practical to researchers.  In the Hub Model, $A_{ij}$ is defined as the probability that node $v_i$ will include node $v_j$ when $v_i$ is the hub node of  a group. The formal definition of the Hub Model will be given in Section \ref{S:Model}.    
	
\begin{figure}[H]
		\centering
		\begin{subfigure}[b]{0.2\textwidth}
			\includegraphics[width=\textwidth]{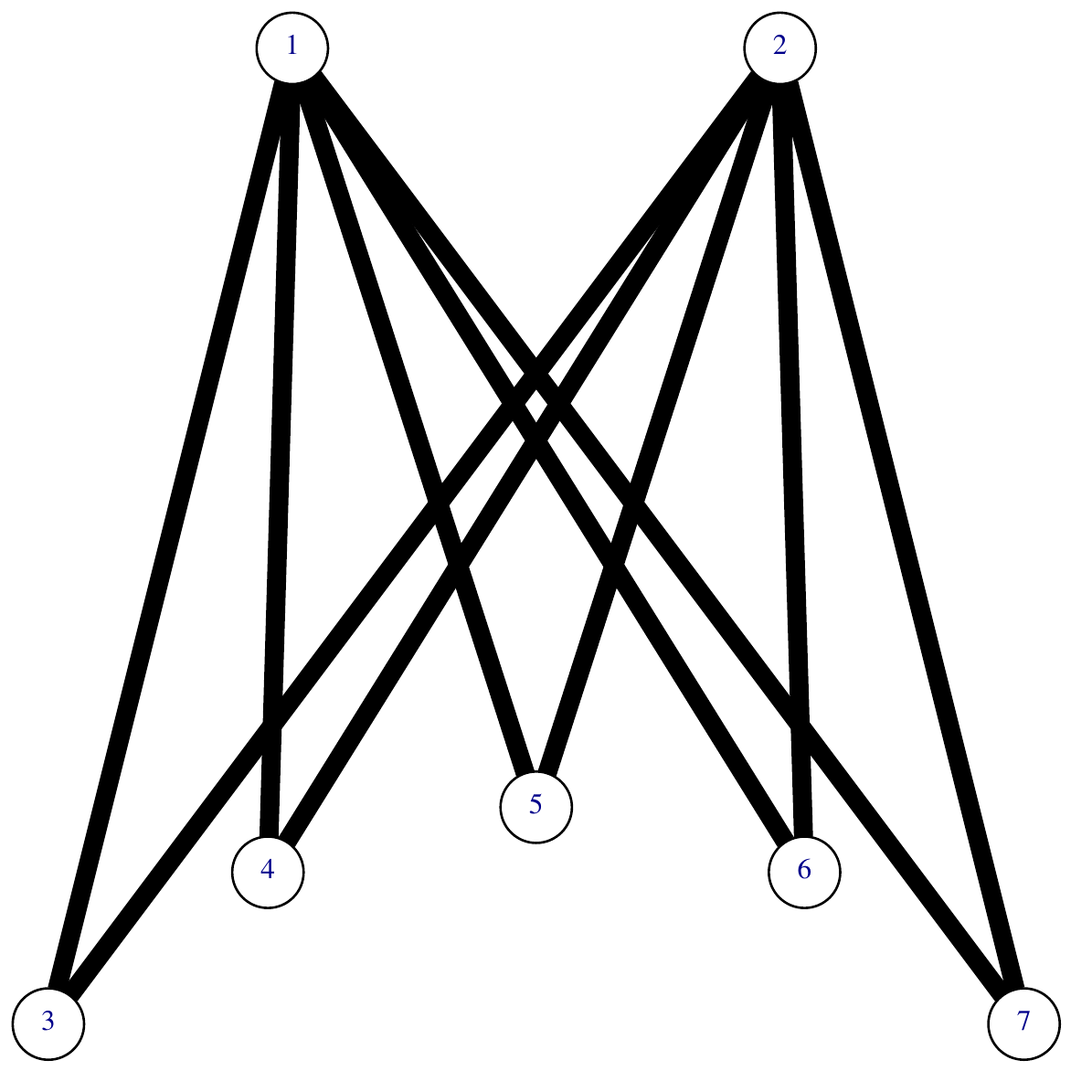}
			\caption{\small Truth}
			\label{F:Shining_True}
		\end{subfigure}%
		~ 
		\begin{subfigure}[b]{0.2\textwidth}
			\includegraphics[width=\textwidth]{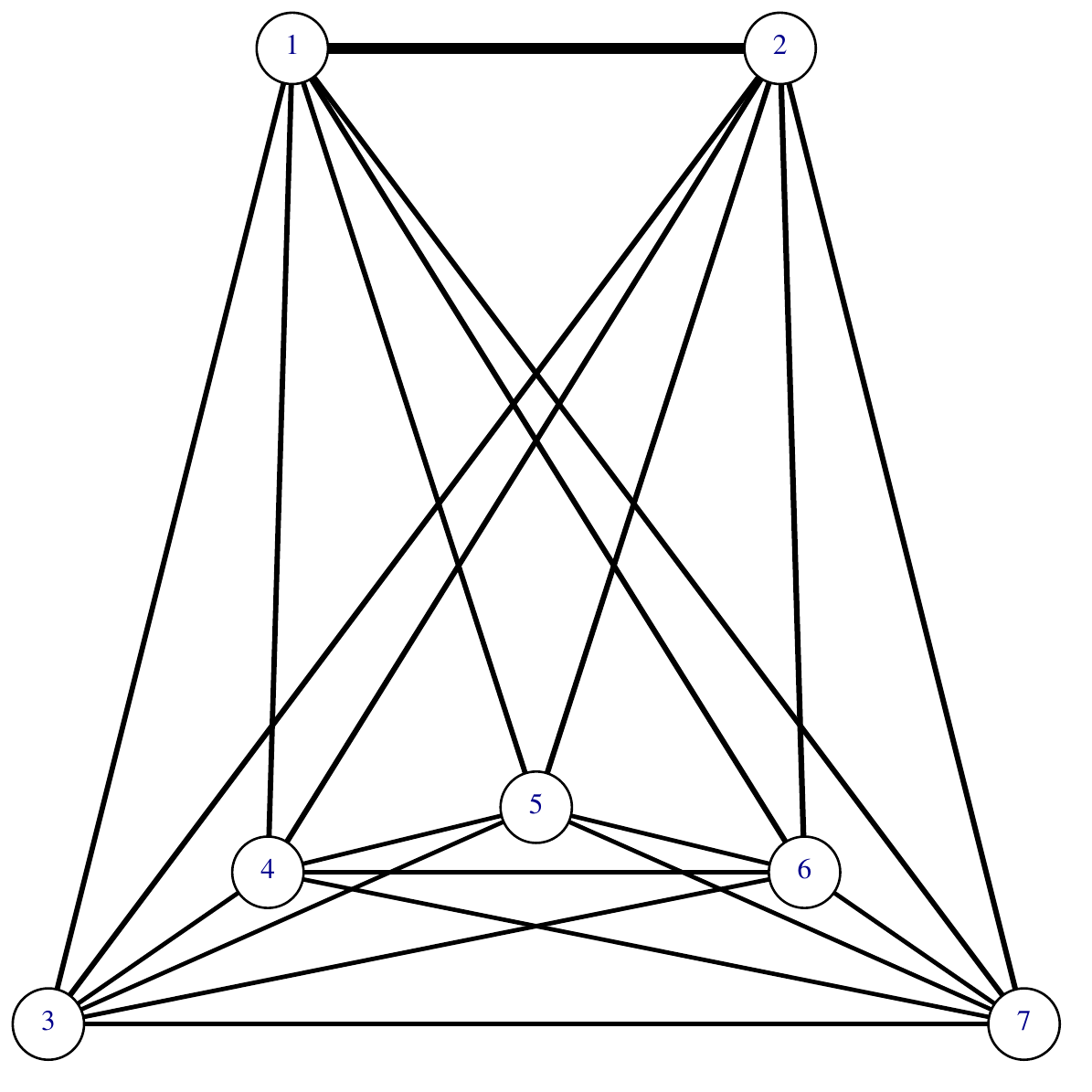}
			\caption{\small Co-occurrence}
			\label{F:Shining_O}
		\end{subfigure}
		~
		\begin{subfigure}[b]{0.2\textwidth}
			\includegraphics[width=\textwidth]{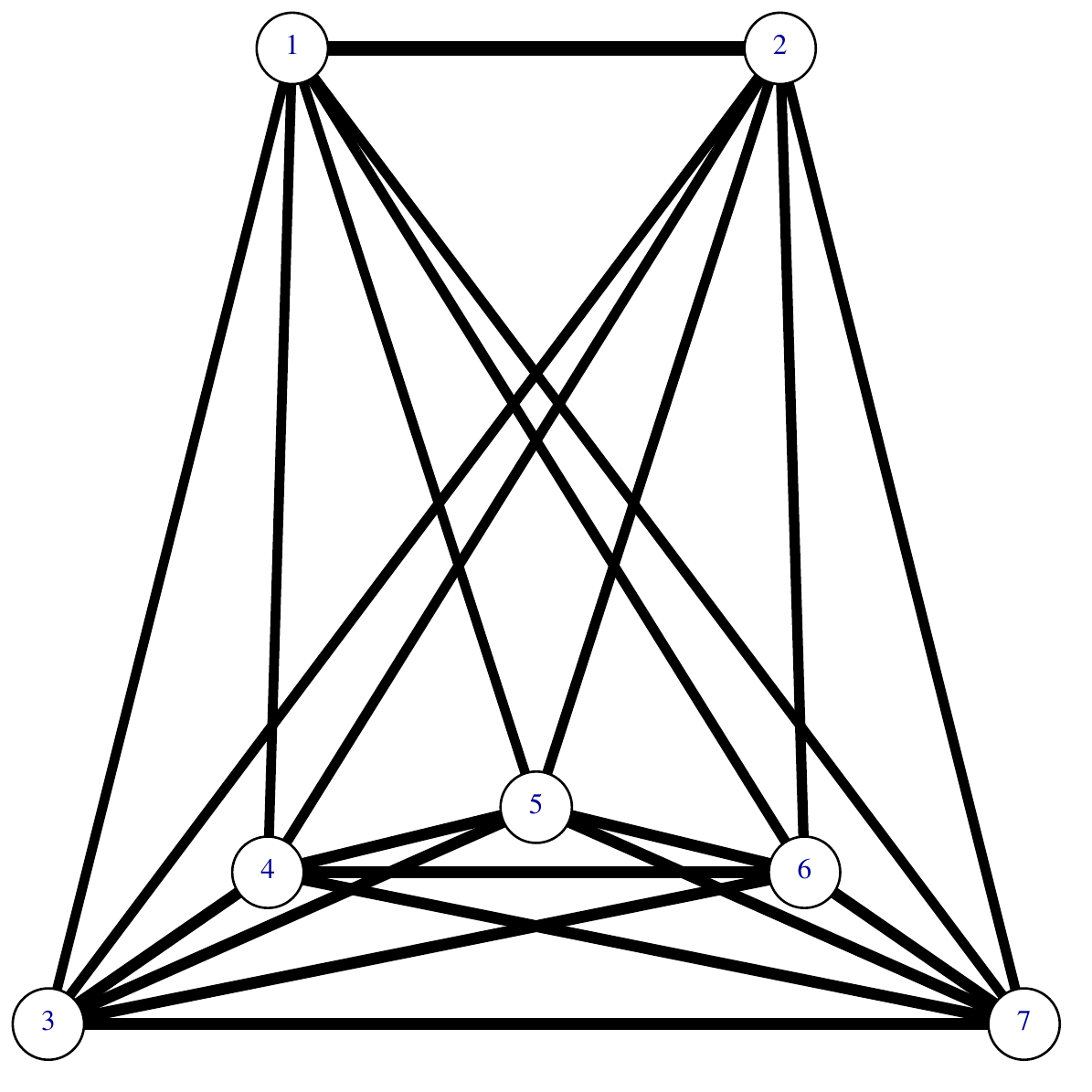}
			\caption{\small Half Weight Index}
			\label{F:Shining_H}
		\end{subfigure}
		~ 
		\begin{subfigure}[b]{0.2\textwidth}
			\includegraphics[width=\textwidth]{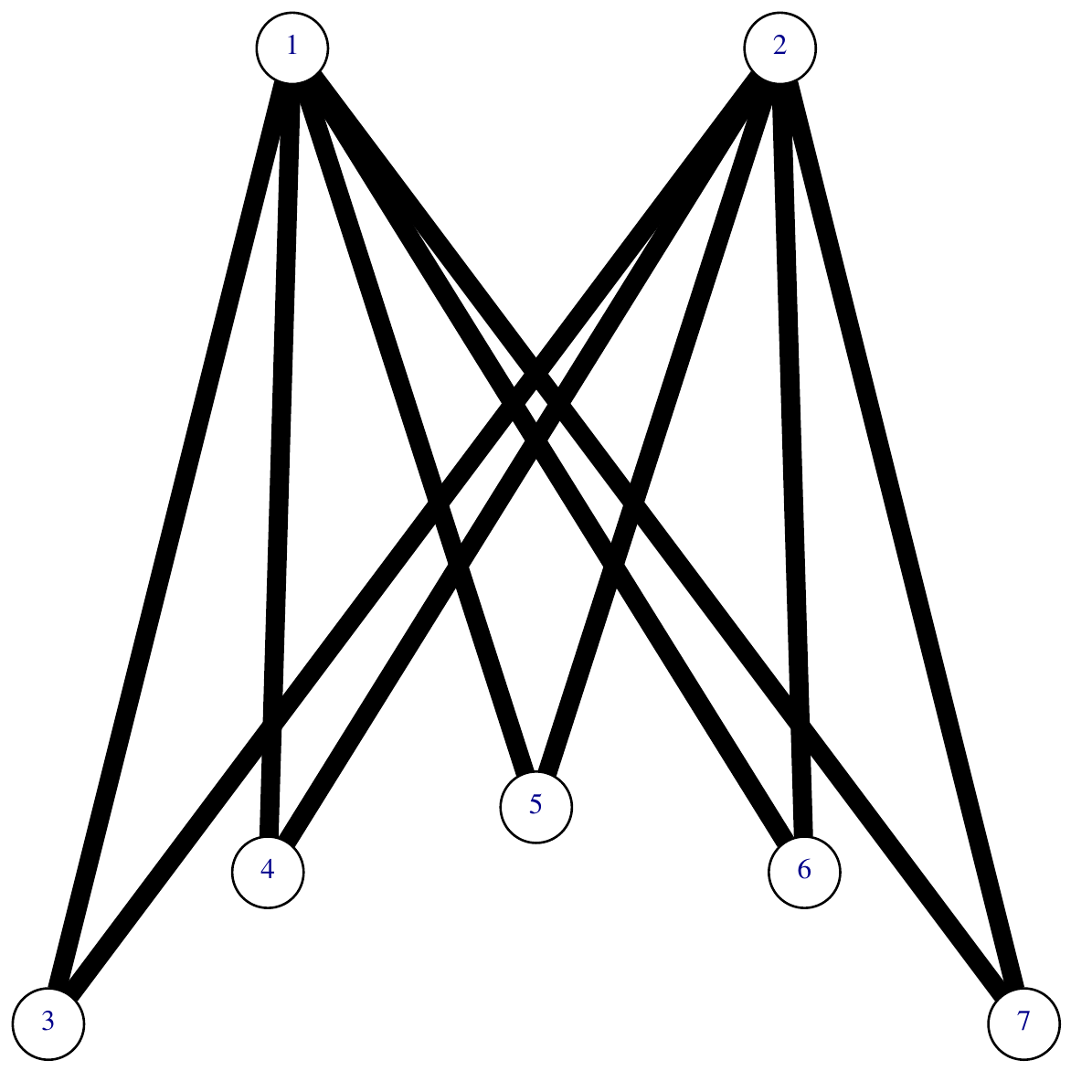}
			\caption{\small Hub Model}
			\label{F:Shining_A}
		\end{subfigure}
		\caption{Comparison of Estimation Techniques}\label{F:Shining}
	\end{figure}

	As a graphical introduction to the performance of Hub Models, consider Figure \ref{F:Shining_True} where relationship strength is represented by the width of the links between nodes.  In this example there is a pair of nodes, $v_1$ and $v_2$, which never directly bond to each other.  Despite the aversion between these two nodes, they share a number of strong relationships with common nodes and therefore often co-occur.  In Figure \ref{F:Shining_O}, we can see that the co-occurrence matrix mistakenly assigns a relatively strong relationship to nodes $v_1$ and $v_2$. In Figure \ref{F:Shining_H}, the half weight index arrives at a very similar conclusion as the co-occurrence matrix.  In both Figures \ref{F:Shining_O} and \ref{F:Shining_H}, the non-existent relationship between nodes $v_1$ and $v_2$ is actually estimated to be stronger than all other relationships. By contrast, results of the Hub Model in Figure \ref{F:Shining_A} clearly capture the latent network of the population.  
	
	To the best of our knowledge, there have been very limited attempts to infer latent networks by model-based approaches. Rabbat et al.\cite{Rabbat08} provides an application for telecommunication networks. Rabbat et al. modeled the process of group formation as a random walk from a source node to a terminal node.  The nodes along the path were subjected to an unknown permutation to account for the lack of order information.  Treating permutations as missing data, Rabbat et al. employed a \textit{Monte Carlo EM} algorithm based on importance sampling to estimate the parameters of the model. This model assumed a distinctly different process of group formation.
	
	In the following sections we present a formal description of the grouped data structure, review existing techniques, and define Hub Models.  Then we study Hub Model identifiability and prove that a symmetry constraint on the adjacency matrix is a sufficient condition for identifiability. We propose an EM algorithm to solve the maximum likelihood estimator of Hub Model.  We then evaluate the performance of the proposed method by simulation studies. We apply the Hub Model to infer the relationships between United States Senators in the 110$^{th}$ Congress based on their co-sponsorship of legislation, the relationships among the characters of the 18$^{th}$ century Chinese novel, \textit{Dream of the Red Chamber}, and apply the Hub Model to floral dispersion over North America.  We close with a discussion that returns to the subject of identifiability along with a phenomenon of ``self-sparsity'' of the Hub Model estimators. 
	
	\section{GROUPED DATA}

	\subsection{Data Structure}
	
	For a population of $n$ individuals, $V=\{v_1,\dots, v_n\}$, we observe  $T$ subsets of the global population, $\{V^{(t)}|V^{(t)} \subseteq V, t=1,...,T \}$.  
	Each observed subset $V^{(t)}$ can be coded as an $n$ length row vector $G^{(t)}$ where:
	
	\[ G_i^{(t)} = \left\{ 
	\begin{array}{l l}
	1 & \quad \textnormal{if $v_i\in V^{(t)}$}\\
	0 & \quad \textnormal{if $v_i \notin V^{(t)}$}
	\end{array} \right.\]
	
	The full set of observations is denoted by a $T \times n$ matrix, $G$.  The $t^{th}$ row of $G$ is $G^{(t)}$.
	
	\subsection{Existing Methods}
	
	Inferring latent networks from grouped data relies on descriptive statistics which count the number of times that two nodes are observed together.  We focus on two popular techniques which estimate probabilities of individual behavior.

	An intuitive and computationally simple measure of grouped data is the \textit{co-occurrence matrix}.  Versions of this technique appear throughout the literature under many names and notations including: \textit{capacity matrix} \cite{Zachary77}, \textit{sociomatrix} \cite{Wasserman94}, \textit{inter-citation frequency} \cite{Kolaczyk09}, \textit{cocitation matrix} \cite{Newman11}, and \textit{strength} \cite{Brent11}.
	
	
	A co-occurrence matrix, $O$, is an $n\times n $ symmetric matrix, defined by:  
	\begin{equation}
	O=\frac{G' G}{T},
	\end{equation}
	
	which estimates the frequency that the nodes $v_i$ and $v_j$ are observed in the same group.
	
	One shortcoming of the co-occurrence matrix is that it estimates the probability that two nodes \textit{will be observed} to co-occur in a given observation.  That is, if two nodes have a strong relationship, but appear in the dataset infrequently, the co-occurrence matrix will estimate a low probability that the two nodes \textit{will be observed} to co-occur.
	
	As an example, consider four nodes $v_1,\dots, v_4$ and the grouped data represented in Table \ref{T:O_vs_H_Example}.
	{\linespread{1.0}
		\begin{table}[hbt!]
			\begin{center}
				\begin{tabular}{|c  | c c c c  |}
					\hline
					
					&\multicolumn{4}{c|}{Node} \\
					\hline
					Event & $v_1$ & $v_2$ & $v_3$ & $v_4$  \\
					\hline
					1 & 1 & 0 & 0 & 0 \\
					2 & 1 & 1 & 0 & 0 \\
					3 & 1 & 1 & 0 & 0 \\
					4 & 1 & 0 & 1 & 1 \\
					5 & 0 & 1 & 1 & 1 \\
					\hline
					
				\end{tabular}
			\end{center}
			\caption{Notional Grouped Data}
			\label{T:O_vs_H_Example}
		\end{table}
	}
	For this dataset, both $O_{1,2}=\frac{2}{5}$ and $O_{3,4}=\frac{2}{5}$.  However, notice that every time node $v_3$ is present node $v_4$ is also present.  A researcher may conclude that there is some aspect of the relationship between nodes $v_3$ and $v_4$ which has been understated.
	
	As an alternative, the \textit{half weight index} estimates the probability that two nodes will be observed to co-occur given that one of them is observed \cite{Cairns87}.
	
	The half weight index has been introduced in a number of equivalent forms \cite{Dice45}.  Computationally, the most direct form is: 
	\begin{equation}\label{E:HWI1}
	H_{ij} = \frac{2\sum_t G_i^{(t)} G_j^{(t)}}{\sum_t G_i^{(t)}+\sum_t G_j^{(t)}}.
	\end{equation}
	
	The half weight index estimates the frequency that the nodes $v_i$ and $v_j$ are observed in the same group given that one of them is observed. 
	
	Returning to the example in Table \ref{T:O_vs_H_Example}, we can see that $H_{1,2}=\frac{4}{7}$ while $H_{3,4}=\frac{4}{4}$.  Therefore, the half weight index provides different information about the latent network than the co-occurrence matrix.
	
	\section{HUB MODELS}\label{S:Model}
	
	\subsection{Generating Mechanism}

	Hub Models (HM) assume that each group is a star subgraph on the global population. The hub node of the star subgraph connecting $G^{(t)}$ is represented by an $n$ length row vector, $S^{(t)}$, where
	
	\[ S_i^{(t)} = \left\{ 
	\begin{array}{l l}
	1 & \quad \text{if $v_i$ the hub node of sample $t$},\\
	0 & \quad \text{otherwise}.
	\end{array} \right.\]
	
	There is one and only one element of $S^{(t)}$ that is equal to 1. 
	
	Under the Hub Model, each group $G^{(t)}$ is independently generated by a two step process. 
	
	\begin{enumerate}
		\item The hub node is drawn from a multinomial distribution with parameter $\rho=(\rho_1,...,\rho_n)$, i.e., $\rho_i=\mathbb{P}(S_i^{(t)}=1)$. The following constraint applies to $\rho_i$:
		\begin{equation}\label{E:USM Constraint}
		\sum_i \rho_i=1.
		\end{equation} 
		
		\item The hub node, $v_i$, chooses to include $v_j$ in the group with probability $A_{ij}$, i.e., $A_{ij}=	\mathbb{P}(G_j^{(t)}=1|S_i^{(t)}=1)$.
	\end{enumerate}
	
	In most practical applications, the hub node of each group $S^{(t)}$ is missing data. We focus on this case in this article. When necessary,we  refer to the model with both $S$ and $G$ being observed as the Known Hub Model (KHM).
	
	Since many existing techniques for inferring networks produce a symmetric adjacency matrix, we assume $A_{ij}=A_{ji}$. The symmetry condition will be shown to ensure the identifiability of the Hub Model when $S$ is unobserved(Section \ref{S:Ident}).
	
	Further, we assume that the hub node will always include itself in the group, i.e. $A_{ii}=1$ for all $i$.
	
	This generating mechanism implies that each observed group is independent of every other observed group.  In particular, $G^{(t)}$ is not a transformation of $G^{(t-1)}$ and the order in which groups are observed contains no information about the relationships between group members.  Researchers often collect data in such a way to ensure this property \cite{Bejder98}. 
	
	\subsection{Likelihood of the Hub Model}
	
	From now on, when we mention the likelihood of the Hub Model, we mean the marginal likelihood of the group observations $G$.
	
	Under the HM, the probability of $G^{(t)}$ has the form of a finite mixture model with $n$ components:
	\begin{equation}\label{E:HM ProbOfGt}
	\mathbb{P}(G^{(t)}|A,\rho)=\sum_{i=1}^n \rho_i G_i^{(t)} \prod_{j}{A_{ij}^{G_j^{(t)}} (1-A_{ij})^{1-G_j^{(t)}}}.
	\end{equation}
	
	By taking the log of the product of individual observed groups, the log likelihood function for the full set of observations is:
	\begin{equation}\label{E:HM LL}	\mathcal{L}(G|A,\rho)=\sum_t\textnormal{log}\Big[\sum_{i=1}^n \rho_i G_i^{(t)} \prod_{j}{A_{ij}^{G_j^{(t)}} (1-A_{ij})^{1-G_j^{(t)}}}\Big].
	\end{equation}
	
	Solving the MLE of HM is an optimization problem with the equality constraints $\sum_i \rho_i=1$, and $A_{ij}=A_{ji}$ for all $i$ and $j$. From \eqref{E:HM LL}, we denote the log likelihood function as $\mathcal{L}(G|A,\rho)$.  This gives the following Lagrange function:
	\begin{equation}\label{E:HM Objective Function}
	\Lambda(G|A,\rho)=\mathcal{L}(G|A,\rho)-\lambda_o  [(\sum_i \rho_i)-1  ]-\sum_{i<j}\lambda_{ij}(A_{ij}-A_{ji}).
	\end{equation}
	
	The log likelihood does not have a closed-form solution for the MLE.  Instead we will derive estimating equations which can be incorporated into an Expectation Maximization algorithm.  Before doing so we investigate the identifiability problem of the Hub Model.
	
	\subsection{Ensuring Identifiability}\label{S:Ident}
	
	A basic requirement for any model is \textit{identifiability}.  For Hub Models, this means for any two sets of parameters $\{ A,\rho \}$ and $\{A^*, \rho^*\}$:
	\begin{align}
	\mathbb{P}(G=g|A,\rho)=\mathbb{P}(G=g|A^*,\rho^*) \hspace{2mm} \forall g \implies A=A^*, \rho=\rho^* .
	\end{align}  
	
	We first show that if no constraint is put on $A$, the model is unidentifiable.
	
	This can be demonstrated by the following simple counterexample.  Consider a network of size $n=4$ with $\{A,\rho\}$ defined in Table \ref{T:IdentEx}. Notice that nodes $v_1$ and $v_2$ both always produce the same group while nodes $v_3$ and $v_4$ produce a different group.
	
	{\linespread{1.0}
		\begin{table}[hbt!]
			\begin{center}
				\begin{tabular}{| l | c | l l l l|}
					\cline{2-6}
					\multicolumn{1}{c}{}  &  \multicolumn{5}{|c|}{$A_{ij}$}\\
					\cline{3-6}
					\multicolumn{1}{c}{}  & \multicolumn{1}{|c}{}    & \multicolumn{4}{|c|}{$j$}\\
					\hline
					$\rho_i$ &  $i$ & 1 & 2 & 3 & 4\\
					\hline
					0.25 & 1 & 1 & 1 & 1 & 0 \\
					0.25 & 2 & 1 & 1 & 1 & 0 \\
					0.25 & 3 & 0 & 1 & 1 & 1 \\
					0.25 & 4 & 0 & 1 & 1 & 1 \\
					\hline                 
				\end{tabular}
			\end{center}
			\caption{Example of a Set of Parameters Which are Not Identifiable}
			\label{T:IdentEx}
		\end{table}
	}
	The probability of $G$ has the form:
	
	\[ \mathbb{P}(G=g|A,\rho) = \left\{ 
	\begin{array}{l l}
	\frac{1}{2} & \quad  g=\{1,1,1,0\}, \\
	\frac{1}{2} & \quad  g=\{0,1,1,1\}, \\
	0 & \textnormal{otherwise} .
	\end{array} \right.\] 
	
	There are an infinite number of parameters yielding the same distribution, but a simple alternative is as follows.  Let $\rho^*=(0.5, 0 , 0.5,0)$, leave the first and third rows of $A$ unchanged but let all other components  of $A^*$ assume arbitrary values. Obviously, we have $	\mathbb{P}(G=g|A,\rho)=\mathbb{P}(G=g|A^*,\rho^*)$ for all $g$. This counterexample demonstrates that the model requires an additional condition to be identifiable.
	
	The following theorem shows that symmetry of $A$ is a sufficient condition for identifiability.
	
	\newtheorem{HM}{Theorem}
	
	\begin{HM}\label{HM}
		Let $A$ and $A^*$ be symmetric adjacency matrices with $A_{ii}=A_{ii}^*=1$ for all $i$, $A_{ij}<1$ and $A_{ij}^*<1$ for all ${i\neq j}$. If  $\mathbb{P}(g|A,\rho)=\mathbb{P}(g|A^*,\rho^*)$ for all $g$, then $\{A,\rho\}=\{A^*,\rho^*\}$. 
	\end{HM}
	
	Let $g^x$ and $g^y$ denote the singleton groups which consist only of nodes $v_x$ and $v_y$, respectively.  Further, let $g^{xy}$ denote the group representing the pair of $v_x$ and $v_y$.
	
	From \eqref{E:HM ProbOfGt} the probability of the singletons is:
	\begin{align}
	\mathbb{P}(g^x|A,\rho)&=\rho_x (1-A_{xy}) \prod_{j\ne\{x,y\}}(1-A_{xj})\label{E:Prob_gx}\\
	\mathbb{P}(g^y|A,\rho)&=\rho_y (1-A_{xy}) \prod_{j\ne\{x,y\}}(1-A_{yj})\label{E:Prob_gy}.
	\end{align}
	
	In \eqref{E:Prob_gy} we have taken advantage of the symmetry of $A$ to replace $A_{yx}$ with $A_{xy}$.
	
	Now, we consider the probability of $g^{xy}$.
	\begin{align}
	\mathbb{P}(g^{xy}|A,\rho)& =\rho_x A_{xy} \prod_{j\ne\{x,y\}}(1-A_{xj})
	+ \rho_y A_{xy} \prod_{j\ne\{x,y\}}(1-A_{yj}) \notag \\
	& =A_{xy} \Big[\rho_x \prod_{j\ne\{x,y\}}(1-A_{xj})
	+ \rho_y  \prod_{j\ne\{x,y\}}(1-A_{yj})\Big]  \notag \\
	& =A_{xy} \Big[ \frac{\mathbb{P}(G=g^x|A,\rho)}{(1-A_{xy})}
	+ \frac{\mathbb{P}(g^y|A,\rho)}{(1-A_{xy})}\Big]  \notag \\
	&	=\frac{A_{xy}}{(1-A_{xy})} \Big[ \mathbb{P}(g^x|A,\rho)
	+ \mathbb{P}(g^y|A,\rho)\Big], \label{E:Prob_gxy}
	\end{align}
	which implies that:
	\begin{equation}\label{E:HM_MoM}
	A_{xy}=\frac{\mathbb{P}(g^{xy}|A,\rho)}{\mathbb{P}(g^x|A,\rho)
		+ \mathbb{P}(g^y|A,\rho)+\mathbb{P}(g^{xy}|A,\rho)}.
	\end{equation}
	
	Therefore, $A_{xy}=A_{xy}^*$ for all $x$ and $y$. 
	
	To complete the proof, consider an arbitrary node $v_x$ which appears as a singleton represented by $g^x$:
	\begin{equation}
	\mathbb{P}(g^x|A,\rho)=\rho_x  \prod_{j \ne x }(1-A_{xj}).
	\end{equation}
	If $A_{xy}=A_{xy}^*$ for all $x$ and $y$ and $\mathbb{P}(g|A,\rho)=\mathbb{P}(g|A^*,\rho^*)$ for all $g$, then:
	\begin{equation}
	\rho_x  \prod_{j \ne x }(1-A_{xj})=\rho_x^*  \prod_{j \ne x }(1-A_{xj})
	\end{equation}
	and it is easy to see that $\rho_x=\rho_x^*$ for all $x$.
	
	$\square$
	
	\paragraph{Remarks}
	
	Before proceeding we would like to make three remarks about Hub Models.
	
	\begin{enumerate}
		\item The generating mechanism for Hub Models is equivalent to a finite mixture model of multivariate Bernoulli random variables. In general, such a model is not identifiable \cite{Teicher61}. This shortcoming does not prevent such models from being useful in many applications. For example, when dealing with classification problems where the researcher only has to identify which component density an observation came from, this type of mixture can be effectively used \cite{Carreira00}.  In such a situation, the individual parameters of the multivariate Bernoulli random variables are not of interest.  However, the issue of identifiability presents a challenge in network inference because we are specifically interested in the individual parameters of $A$. 
		
		\item Equation \eqref{E:HM_MoM} suggests a method of moments estimator for $A_{xy}$ based on frequencies of doubletons and singletons. However,  this estimator requires that the probability of doubletons and singletons be estimated accurately so this technique would be very inefficient for most real cases, because small groups appear infrequently in many datasets. Therefore, we will continue to consider the MLE which presumably uses all available information. 
		\item It is worth noticing that even though symmetry of $A$ is a natural assumption in network analysis, it is only a sufficient condition for identifiability according to Theorem \ref{HM}. For future work, we will explore other assumptions to ensure identifiability and ultimately find a necessary and sufficient condition. 
		
	\end{enumerate} 
	
	\subsection{Estimating Equations}
	
	The maximum likelihood estimator of HM does not have a closed-form solution for the parameters. In this section, we derive estimating equations for the conditions that the MLE $\{\hat{A}, \hat{\rho}\}$ must satisfy. Then we will show that solving these equations iteratively is equivalent to an EM algorithm. The details of the EM algorithm will be given in the next section. 
	
	We begin by taking the derivative of \eqref{E:HM LL} with respect to $A_{xy}$ and $A_{yx}$.
	
	\begin{align}
	\frac{\partial \Lambda(G|A,\rho)}{\partial A_{xy}}& =\frac{\partial \mathcal{L}(G|A,\rho)}{\partial A_{xy}}-\lambda_{xy}=0 \mbox{ if } x<y, \\
	\frac{\partial \Lambda(G|A,\rho)}{\partial A_{yx}}& =\frac{\partial \mathcal{L}(G|A,\rho)}{\partial A_{yx}}+\lambda_{xy}=0 \mbox{ if } x>y.
	\end{align}
	Therefore, 
	\begin{equation}\label{E:Symmetric Constraint}
	\frac{\partial \mathcal{L}(G|A,\rho)}{\partial A_{xy}}=-\frac{\partial \mathcal{L}(G|A,\rho)}{\partial A_{yx}}.
	\end{equation}
	
	We now focus on the derivative of the log likelihood function of \eqref{E:HM LL}:
	\begin{equation}\label{E:HM Derivative of LL1}
	\sum_t \frac{\rho_x G_x^{(t)}\Big[\prod_{j \ne y}{A_{xj}^{G_j^{(t)}} (1-A_{xj})^{1-G_j^{(t)}}}\Big]\frac{\partial}{\partial A_{xy}} \Big( A_{xy}^{G_y^{(t)}}(1-A_{xy})^{1-G_y^{(t)}} \Big) }{\sum_{i=1}^n \rho_i G_i^{(t)} \prod_{j}{A_{ij}^{G_j^{(t)}} (1-A_{ij})^{1-G_j^{(t)}}}}.
	\end{equation}
	
	Note that the derivative in the numerator of \eqref{E:HM Derivative of LL1} is equal to 1 if node $v_y$ is in observation $G^{(t)}$, and $-1$ if $v_y$ is not in the observation.  We represent this by the function:
	
	\[ \gamma(G_y^{(t)}) = \left\{ 
	\begin{array}{l l}
	1 & \quad \textnormal{if $G_y^{(t)}=1$},\\
	-1 & \quad \textnormal{if $G_y^{(t)}=0$}.
	\end{array} \right.\]
	Therefore, 
	\begin{equation}\label{E:HM Derivative of LL2}
	\frac{\partial}{\partial A_{xy}} \mathcal{L}(G|A,\rho)=\sum_t  \frac{\Big[ \rho_x G_x^{(t)}\prod_{j \ne y}{A_{xj}^{G_j^{(t)}} (1-A_{xj})^{1-G_j^{(t)}}}\Big]\gamma(G_y^{(t)})}{\sum_{i=1}^n \rho_i G_i^{(t)} \prod_{j}{A_{ij}^{G_j^{(t)}} (1-A_{ij})^{1-G_j^{(t)}}}}.
	\end{equation}
	
	The denominator of \eqref{E:HM Derivative of LL2} is simply the probability of $G^{(t)}$ (see \eqref{E:HM ProbOfGt}).  In addition, the term in brackets can be made equal to $\mathbb{P}(G^{(t)},S_x=1)$ by multiplying $A_{xy}^{G_y^{(t)}}(1-A_{xy})^{(1-G_y^{(t)})}$. To conserve space, we suppress $\{A,\rho\}$ going forward.  This gives:
	\begin{equation}\label{E:HM Derivative of LL3}
	\frac{\partial}{\partial A_{xy}} \mathcal{L}(G)=\sum_t \frac{\gamma(G_y^{(t)}) \mathbb{P}(G^{(t)}, S_x^{(t)}=1)}{A_{xy}^{G_y^{(t)}}(1-A_{xy})^{(1-G_y^{(t)})} \mathbb{P}(G^{(t)})}.
	\end{equation}
	
	This equation can be further simplified by observing that  $\frac{\mathbb{P}(G^{(t)},S_x^{(t)}=1)}{\mathbb{P}(G^{(t)})}$ is equivalent to $\mathbb{P}(S_x^{(t)}=1|G^{(t)})$:
	\begin{equation}\label{E:HM Derivative of LL Final}
	\frac{\partial}{\partial A_{xy}} \mathcal{L}(G)=\sum_t \frac{\gamma(G_y^{(t)}) \mathbb{P}(S_x^{(t)}=1|G^{(t)})}{A_{xy}^{G_y^{(t)}}(1-A_{xy})^{(1-G_y^{(t)})}}.
	\end{equation}
	
	Plugging \eqref{E:HM Derivative of LL Final} into \eqref{E:Symmetric Constraint}, we get:
	\begin{equation}
	\sum_t \frac{\gamma(G_y^{(t)}) \mathbb{P}(S_x^{(t)}=1|G^{(t)})}{A_{xy}^{G_y^{(t)}}(1-A_{xy})^{(1-G_y^{(t)})}}= 
	- \sum_t \frac{\gamma(G_x^{(t)}) \mathbb{P}(S_y^{(t)}=1|G^{(t)})}{A_{yx}^{G_x^{(t)}}(1-A_{yx})^{(1-G_x^{(t)})}}.
	\end{equation}
	
	By applying symmetry and breaking the summations, this becomes:
	\begin{align}
	\sum_{t:G_y^{(t)}=1} \frac{\mathbb{P}(S_x^{(t)}=1|G^{(t)})}{A_{xy}}
	&- \sum_{t:G_y^{(t)}=0} \frac{\mathbb{P}(S_x^{(t)}=1|G^{(t)})}{1-A_{xy}}\notag\\
	= - \sum_{t:G_x^{(t)}=1} \frac{\mathbb{P}(S_y^{(t)}=1|G^{(t)})}{A_{xy}}
	&+ \sum_{t:G_x^{(t)}=0} \frac{\mathbb{P}(S_y^{(t)}=1|G^{(t)})}{1-A_{xy}}.
	\end{align}
	With some simple algebra, it is easy to see that:
	\begin{equation}\label{E:HM A_hat}
	\hat{A}_{xy}=\frac{\sum_t G_y^{(t)} \mathbb{P}(S_x=1|G^{(t)})+\sum_t G_x^{(t)}\mathbb{P}(S_y=1|G^{(t)})}{\sum_t \big[\mathbb{P}(S_x=1|G^{(t)})+\mathbb{P}(S_y=1|G^{(t)})\big]}.
	\end{equation}
	
	It is worth repeating that \eqref{E:HM A_hat} is not a closed form solution for $\hat{A}_{xy}$.  This is because the right hand side of the equation depends on $\hat{A}_{xy}$. 
	
	We next derive the estimating equation for $\hat{\rho}$. 
	By taking the derivative of \eqref{E:HM Objective Function} with respect to $\rho_x$, we get the following:
	\begin{align}
	\frac{\partial}{\partial \rho_x} \Lambda(G)&=\sum_t \frac{ G_x^{(t)} \prod_{j}{A_{xj}^{G_j^{(t)}} (1-A_{xj})^{1-G_j^{(t)}}}}{\mathbb{P}(G^{(t)})}-\lambda_o\notag\\
	&=\sum_t \frac{ \mathbb{P}(G^{(t)},S_x^{(t)}=1)}{\rho_x \mathbb{P}(G^{(t)})}-\lambda_o\notag\\
	&=\frac{1}{\rho_x}\sum_t \mathbb{P}(S_x^{(t)}=1|G^{(t)})-\lambda_o.
	\end{align}
	Solving this equation for zero, we obtain:
	\begin{equation}\label{E:HM rho with lambda}
	\rho_x=\frac{1}{\lambda_o}\sum_t \mathbb{P}(S_x^{(t)}=1|G^{(t)}).
	\end{equation}
	Using the constraint on $\rho$ \eqref{E:USM Constraint}, we get:
	\begin{equation}\label{E:HM rho_hat}
	\hat{\rho}_x=\frac{\sum_{t=1}^T \mathbb{P}(S_x^{(t)}=1|G^{(t)})}{T}.
	\end{equation}
	
	\section{EM ALGORITHM}
	
	The previous section derived estimating equations which depended on the probability $\mathbb{P}(S_x^{(t)}=1|G^{(t)})$. This implies a fairly intuitive algorithm updating $\{\hat{A},\hat{\rho}\}$ and $\mathbb{P}(S_x^{(t)}=1|G^{(t)}) $ iteratively, which can be fitted into the general framework of EM algorithm.
	
	The key technique of any EM algorithm is to formulate a complete data model then solve the model as if some data is observed and other data is missing.  In this case, the Known Hub Model serves as the complete data model, $G$ is the observed data, and $S$ is the missing data. Each iteration of the EM algorithm consists of an expectation step followed by a maximization step \cite{McLachlan08}.  
	
	\vspace{4mm}
	\textbf{E-Step}
	\vspace{4mm}
	
	Since the log likelihood function of the complete data model is linear in the unobserved data, $S_i^{(t)}$, the E-Step (on the $(m+1)^{th}$ iteration) simply requires calculating the current conditional expectation of $S_i^{(t)}$ given the observed data, $G^{(t)}$ (see \cite{McLachlan08} for detailed explanation).
	\begin{align}\label{E:ProbKisCenter}
	E[S_x^{(t)}|G^{(t)}]&=\mathbb{P}(S_x^{(t)}=1|G^{(t)})\notag\\
	&=
	\frac{ \rho_x G_x^{(t)} \prod_{j}{A_{xj}^{G_j^{(t)}} (1-A_{xj})^{1-G_j^{(t)}}}}{\sum_{i=1}^n \rho_i G_i^{(t)} \prod_{j}{A_{ij}^{G_j^{(t)}} (1-A_{ij})^{1-G_j^{(t)}}}}
	\end{align}
	
	\vspace{4mm}
	\textbf{M-Step}
	\vspace{4mm}
	
	The M-Step replaces $\mathbb{P}(S_x^{(t)}=1|G^{(t)})$ on the right hand side of \eqref{E:HM A_hat} and \eqref{E:HM rho_hat} with $E[S_x^{(t)}|G^{(t)}]$ from \eqref{E:ProbKisCenter}.
	
	\vspace{4mm}
	%
	
	\vspace{4mm}
	\textbf{Algorithm}
	\vspace{4mm}
	
	Algorithm \ref{A:EM Algorithm} illustrates the details of the algorithm for the Hub Model.  
	
	Several standard techniques are used to improve the performance of the EM algorithm.  Firstly, we run the EM algorithm 10 times with different staring points and choose the solution with the highest likelihood. Secondly, we limit the number of iterations applied to a starting point. This second treatment is based in part on the observation that when this algorithm has a bad starting point, it will take a very long time to converge and the point that it converges to is not close to the maximum.  As a final step, we treat any $\hat{A}_{xy}\le 10^{-4}$ as $\hat{A}_{xy}=0$.  We apply this finishing step to remove clutter from the returned solutions.
	
	\begin{spacing}{0.6}
		\begin{algorithm}[H]
			\KwData{G}
			\KwResult{$\hat{A}, \hat{\rho}$}
			Initialize:\\
			$\mathcal{L}(G|\hat{A})=-\infty$\\
			\For{rep=1 to 10}{
				Initialize:\\
				$\hat{A}_{ij}^{(0)}=unif(0,1) \hspace{2 mm} \forall \{i,j\}$\\
				$X_i=unif(0,1) \hspace{2 mm} \forall i$\\
				$\hat{\rho}_i^{(0)}=\frac{X_i}{\sum_k X_k}$\\
				$\Delta \mathcal{L}(G|A^{(0)})=10^4$\\
				counter=1\\
				\While{$\vert \frac{\Delta \mathcal{L}(G|A^{(m+1)})}{ \mathcal{L}(G|A^{(m)})} \vert >10^{-4} \textnormal{and counter}<100$}{
					\textbf{E-Step}\\
					$\>$ Update $\mathbb{P}(S_k^{(t)}=1|G^{(t)})$ by Equation \ref{E:ProbKisCenter}\\
					\textbf{M-Step}\\
					$\>$ Update $A^{(m+1)}$ by Equation \ref{E:HM A_hat}\\
					$\>$ Update $\rho^{(m+1)}$ by Equation \ref{E:HM rho_hat}\\
					$\Delta \mathcal{L}(G|A^{(m+1)})=\mathcal{L}(G|A^{(m+1)})-\mathcal{L}(G|A^{(m)})$\\
					counter=counter$+1$\\
				}
				\If{$\mathcal{L}(G|A^{(m+1)})>\mathcal{L}(G|\hat{A})$}{
					\eIf{$\hat{A}_{ij} \le 10^{-4}$}{
						$\hat{A}_{ij}=0$
					}
					{$\hat{A}_{ij}=A_{ij}^{(m+1)}$}
				}
			}
			\caption{Expectation Maximization Algorithm for the Hub Model}
			\label{A:EM Algorithm}
		\end{algorithm}
	\end{spacing}

	\section{SIMULATION}\label{S:Simulation}

	In order to perform simulations, we generate $\{A,\rho\}$ using the following techniques. 
	
	For $\rho$, we select $n$ i.i.d. random numbers uniformly, $X_i$, and divide each random number by the sum of all $X_i$'s.  That is, $\rho_i=\frac{X_i}{\sum_i X_i}$. 
	
	We use a two step process to generate $A$.  First, we create a symmetric unweighted undirected random graph on $n$ nodes using the configuration model with a power law degree distribution.  We refer to this unweighted graph as the \textit{structure} of the network.  Each edge in the graph is then assigned a relationship strength with a beta distribution,
	\[ A_{ij} = \left\{ 
	\begin{array}{l l}
	Beta(\alpha, \beta) & \textnormal{if there is an edge between $v_i$ and $v_j$}\\
	0 & \textnormal{otherwise} 
	\end{array} \right.\] 
	We simply let $A_{ji}=A_{ij}$ since we assume $A$ is symmetric.
	
	In Table \ref{T:Converge}, we consider four different networks with $n=\{10,20,30,50\}$.  The latent network for each scenario was generated with power $2$, and beta distribution $\alpha=1$, and $\beta=4$. For each network, we generated 100 datasets with $T=50,000$ and estimated the latent network using subsets of the dataset ranging from 100 observations to the full dataset.
	
	For each combination of $n$ and $T$, we report four different measures of performance.  
	
	The first performance measure is the average run time to estimate $\{A,\rho\}$ on a Intel Pentium CPU G2030 at 3.00 GHz with 4.00GB of RAM.
	
	Next, we measured the ability of $\hat{A}$ to correctly identify the structure of the latent network.  To do this we define true positives and true negatives as follows: 
	\begin{align}
	TP_{ij} &=I(  A_{ij} =1 \mbox{ and } \hat{A}_{ij} =1) \\	
	TN_{ij} &=I ( A_{ij} =0 \mbox{ and } \hat{A}_{ij} =0).
	\end{align}
	
	Recall that we apply a threshold to $\hat{A}_{ij}$ in the EM algorithm such that $\hat{A_{ij}}$ is zero if it is less than $10^{-4}$.
	
	Using this notation, we assess the accuracy of the estimated network structure using:
	\begin{equation}
	Accuracy = \frac{\sum_{i<j} TP_{ij}+TN_{ij}}{\binom{n}{2}}
	\end{equation}
	
	\cite{Han11}.
	
	To measure the difference between $\{A,\rho\}$ and $\{\hat{A},\hat{\rho}\}$, we calculate the mean absolute error (MAE):
	\begin{align}
	MAE(A)&=\frac{1}{\binom{n}{2}} \sum_{i<j} |\hat{A}_{ij}-A_{ij}|\\
	MAE(\rho)&=\frac{1}{n} \sum_{i} |\hat{\rho}_{i}-\rho_{i}|.
	\end{align}
	
	The first observation from Table \ref{T:Converge} is simply that as the number of observations increases, the average error in the estimates tends to decline and the variability of the estimate also decreases.
	
	A more interesting observation concerns the runtime of the estimates.  For increasing values of $n$, the average runtime for a single estimate also increases.  However, for all but the case of $n=50$, as the number of observations increases, the run time actually decreases.  This is  due to reductions in the number of iterations necessary to achieve convergence as the dataset gets larger.
	
	The most interesting observation is that while the number of parameters is of order $O(n^2)$ the accuracy of the estimates is high even when there is very little data available.  This is due to a property of the model which we call \textit{self-sparsity} and discuss in more detail in Section \ref{S:Discussion}.  
	
	{\linespread{1.0}
		\begin{table}[H]
			\begin{center}
				\resizebox{5in}{!}{
					\begin{tabular}{|c | c c | c c | c c |}
						\hline
						& \multicolumn{6}{c|}{$n=10$}\\
						\hline
						& Avg Run& Avg & Avg & StDev & Avg & StDev \\
						Obs &  Time (sec) & Accuracy & MAE($\rho$) & MAE($\rho$) & MAE($A$) & MAE($A$)\\
						\hline
						100   & 0.0953 & 0.9329 & 0.0254 & 0.0073 & 0.0132 &  0.0035\\
						200   & 0.0901 & 0.9656 & 0.0181 & 0.0051 & 0.0089 &  0.0022 \\
						500   & 0.0624 & 0.9933 & 0.0115 & 0.0030 & 0.0056 &  0.0015\\
						1000  & 0.0573 & 0.9998 & 0.0081 & 0.0021 & 0.0039 &  0.0011 \\
						2000  & 0.0542 & 1.0000 & 0.0057 & 0.0014 & 0.0027 &  0.0007\\
						5000  & 0.0504 & 1.0000 & 0.0036 & 0.0009 & 0.0017 &  0.0004 \\
						10000 & 0.0494 & 1.0000 & 0.0025 & 0.0007 & 0.0012 &  0.0003\\
						20000 & 0.0500 & 1.0000 & 0.0018 & 0.0005 & 0.0008 &  0.0002\\
						50000 & 0.0504 & 1.0000 & 0.0011 & 0.0003 & 0.0005 &  0.0001\\
						\hline
					\end{tabular}
				}
				\vspace{2mm}
				\resizebox{5in}{!}{
					\begin{tabular}{|c | c c | c c | c c | }
						\hline
						& \multicolumn{6}{c|}{$n=20$} \\
						\hline
						& Avg Run& Avg & Avg & StDev & Avg & StDev \\
						Obs &  Time (sec) & Accuracy & MAE($\rho$) & MAE($\rho$) & MAE($A$) & MAE($A$) \\
						\hline
						100   & 0.2190 & 0.9468 & 0.0194 & 0.0031 & 0.0157 &  0.0054\\
						200   & 0.2121 & 0.9635 & 0.0135 & 0.0024 & 0.0103 &  0.0038\\
						500   & 0.1766 & 0.9756 & 0.0079 & 0.0016 & 0.0048 &  0.0013\\
						1000  & 0.1557 & 0.9861 & 0.0055 & 0.0010 & 0.0030 &  0.0006\\
						2000  & 0.1400 & 0.9945 & 0.0040 & 0.0007 & 0.0021 &  0.0004\\
						5000  & 0.1445 & 0.9997 & 0.0026 & 0.0004 & 0.0013 &  0.0002\\
						10000 & 0.1464 & 1.0000 & 0.0017 & 0.0003 & 0.0009 &  0.0001\\
						20000 & 0.1898 & 1.0000 & 0.0012 & 0.0002 & 0.0006 &  0.0001\\
						50000 & 0.2166 & 1.0000 & 0.0008 & 0.0001 & 0.0004 &  0.000\\
						\hline
					\end{tabular}
				}
				\vspace{2mm}
				
				\resizebox{5in}{!}{
					\begin{tabular}{|c | c c | c c | c c |}
						\hline
						& \multicolumn{6}{c|}{$n=30$} \\
						\hline
						& Avg Run& Avg & Avg & StDev & Avg & StDev \\
						Obs &  Time (sec) & Accuracy & MAE($\rho$) & MAE($\rho$) & MAE($A$) & MAE($A$) \\
						\hline
						
						100   & 0.4226 & 0.9577 & 0.0153 & 0.0021 & 0.0128 &  0.0025 \\
						200   & 0.4306 & 0.9720 & 0.0109 & 0.0017 & 0.0088 &  0.0020 \\
						500   & 0.3419 & 0.9884 & 0.0066 & 0.0010 & 0.0045 &  0.0006 \\
						1000  & 0.3003 & 0.9943 & 0.0047 & 0.0007 & 0.0030 &  0.0004 \\
						2000  & 0.2479 & 0.9960 & 0.0033 & 0.0005 & 0.0021 &  0.0003 \\
						5000  & 0.2286 & 0.9981 & 0.0021 & 0.0004 & 0.0013 &  0.0002 \\
						10000 & 0.2326 & 0.9994 & 0.0014 & 0.0002 & 0.0009 &  0.0001 \\
						20000 & 0.2440 & 0.9999 & 0.0010 & 0.0002 & 0.0006 &  0.0001 \\
						50000 & 0.2481 & 1.0000 & 0.0007 & 0.0001 & 0.0004 &  0.0001 \\
						
						\hline
					\end{tabular}
				}
				\vspace{2mm}
				\resizebox{5in}{!}{
					\begin{tabular}{|c | c c | c c | c c | }
						\hline
						& \multicolumn{6}{c|}{$n=50$}\\
						\hline
						& Avg Run& Avg & Avg & StDev & Avg & StDev \\
						Obs &  Time (sec) & Accuracy & MAE($\rho$) & MAE($\rho$) & MAE($A$) & MAE($A$) \\
						\hline
						
						100   & 0.7410 & 0.9457 & 0.0121 & 0.0012 & 0.0192 &  0.0029\\
						200   & 0.9810 & 0.9562 & 0.0089 & 0.0011 & 0.0142 &  0.0025\\
						500   & 1.2460 & 0.9760 & 0.0054 & 0.0006 & 0.0069 &  0.0014\\
						1000  & 1.2914 & 0.9828 & 0.0037 & 0.0004 & 0.0040 &  0.0004\\
						2000  & 1.3745 & 0.9838 & 0.0025 & 0.0003 & 0.0026 &  0.0002\\
						5000  & 1.8633 & 0.9850 & 0.0016 & 0.0002 & 0.0016 &  0.0001\\
						10000 & 2.5726 & 0.9874 & 0.0011 & 0.0001 & 0.0011 &  0.0001\\
						20000 & 10.9601 & 0.9941 & 0.0008 & 0.0001 & 0.0008 &  0.0001\\
						50000 & 21.0469 & 0.9976 & 0.0005 & 0.0001 & 0.0005 &  0.0000\\
						
						\hline
						
					\end{tabular}
				}
				
			\end{center}
			\caption{Average and Standard Deviation of Mean Absolute Error as Observations Increase}
			\label{T:Converge}
		\end{table}
	}
	As $n$ increases for a fixed number of observations, the estimates of $\rho$ appear to become more accurate. This counterintuitive result is a consequence of the way that $\rho$ is distributed in this set of examples.  As $n$ increases, the mean and variance of $\rho_i$ decreases and leads to the perception of less error.
	
\begin{figure}[H]
		\centering
		\includegraphics[width=3in]{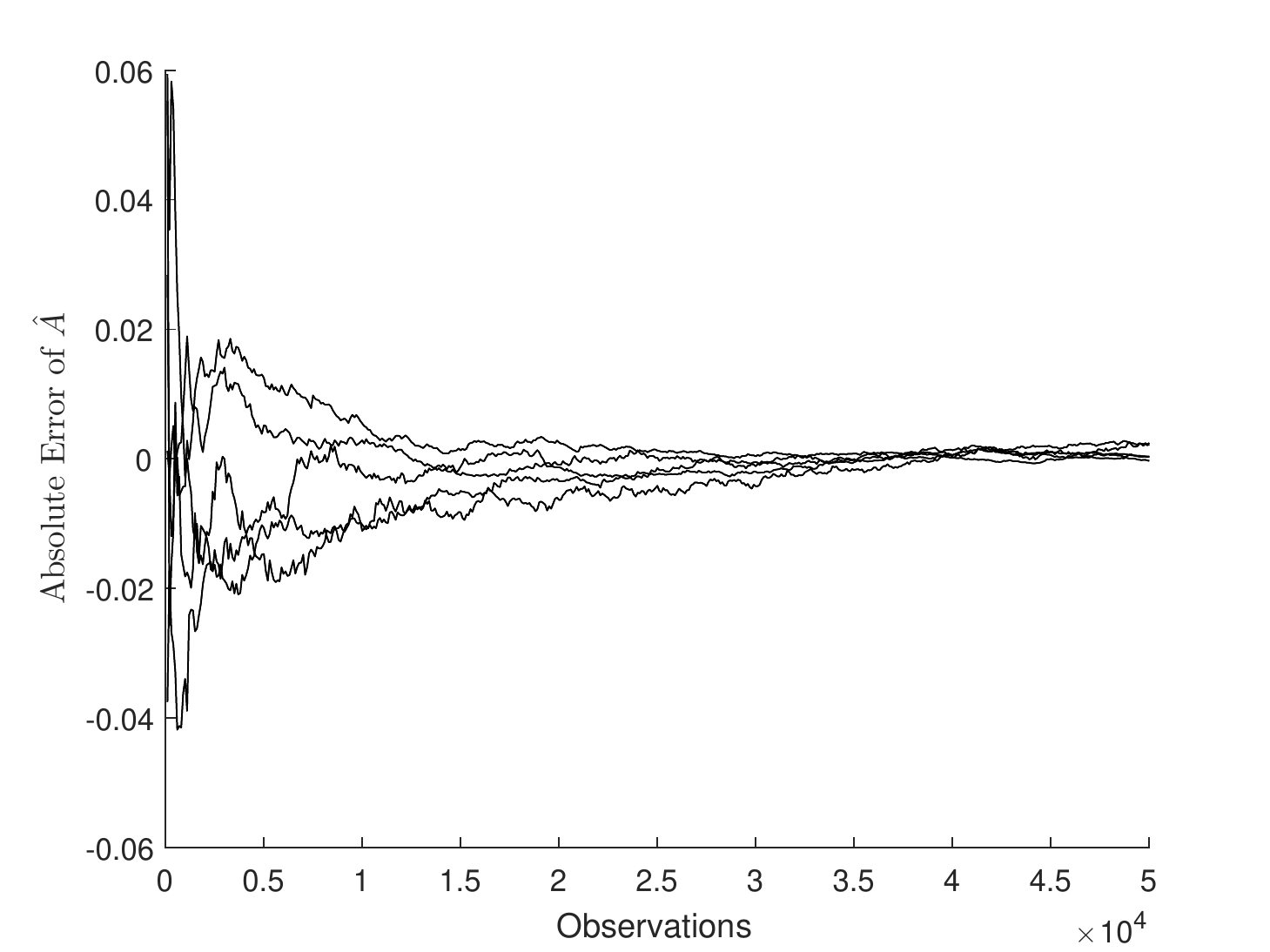}
		\caption{Convergence of Individual Node Estimates}
		\label{F:NodeConverge}
	\end{figure}

	Mean absolute error of the matrix is only a measure of overall estimator performance and cannot give the details of each element of the adjacency matrix.  To explore the manner in which individual elements of the estimate change as the number of observations increases, we use the same latent network with 10 nodes which was used above and generate datasets of increasing size. It can be seen from Figure \ref{F:NodeConverge} that the difference between the parameter and estimates is quickly reduced to very small values and appears to converge to 0.
	
	\section{DATA ANALYSIS}\label{S:Data}

	\subsection{Introduction}
	
	In this section, we perform data analysis on three datasets.
	
	The first dataset records co-sponsorship of legislation in the Senate of the 110$^{th}$ United States Congress.  The rules of the Senate require that each piece of legislation have a unique sponsor; however, other members may co-sponsor the bill \cite{Fowler06a}.  These rules mean that the data conform to the assumption of the Hub Model.
	
	The second dataset is from the 18$^{th}$ century Chinese novel, \textit{Dream of the Red Chamber}.  The observed groups in this dataset do not necessarily conform to the Hub Model assumption.  However, we will show that even without this assumption being explicitly valid, important information about the relationships between individuals can be estimated.
	
	The final dataset has been extracted from the USDA plant database.  Unlike the first two datasets, this one does not deal with ``social'' data, but with ``spatial'' data.  Each observation represents a single species of plant along with each North American state or territory in which the plant is observed to grow.  For this analysis, states and territories represent nodes. As with the second dataset, we find that the Hub Models return meaningful information about underlying structure.
	
	\subsection{Senate of the 110$^{th}$ United States Congress}\label{S:Senate}
	
	The United States Senate is a chamber in the bicameral legislature of the United States, and together with the U.S. House of Representatives makes up the U.S. Congress.  A key function of both chambers of Congress is to originate legislation.  Each piece of legislation can have only one originating sponsor; however, since the mid-1930s, Senators have had an opportunity to express support for a piece of legislation by signing it as a co-sponsor \cite{Fowler06a}.
	
	The 110$^{th}$ United States Congress occurred between January 3, 2007 and January 3, 2009.  The Democratic Party controlled a majority in both chambers for the first time since 1995 with a voting share of 50.5 \% of the Senate membership. 
	
	The United States Senate consists of 100 members with each state represented by 2 Senators at any time.  During this session of Congress, there were a total of 102 individuals who served in the Senate.  One original member died and a second resigned to become a lobbyist.  Both members were replaced by appointed state representatives.
	
	Data for legislative co-sponsorship are available in the Library of Congress Thomas legislative database. This database includes more than 280,000 pieces of legislation proposed in the U.S. House and Senate with over 2.1 million co-sponsorship signatures. Most bills do not pass, and cosponsors need not invest time and resources crafting legislation; so co-sponsorship is a relatively costless way to signal one's position on issues important to constituents and fellow legislators. For the purposes of this study, we include all forms of legislation including all available resolutions, public and private bills, and amendments \cite{Fowler06b}.  During the 110$^{th}$ Congress, the Senate initiated 10,327 pieces of legislation.
	
	It is a trivial task to apply the KHM to this dataset when we treat the sponsor as known; therefore, we focus on the case where the sponsor is unknown.  That is, we intentionally confound sponsors with co-sponsors so that the only data that we have is $G$. We would like to investigate whether the HM can provide a meaningful estimate of the latent network even when the information of hub nodes is missing. The average difference between edges estimated by KHM and HM is 0.03, which suggests that the HM estimate is very accurate even when we confound the hub nodes. 
	
\begin{figure}[H]
		\centering
		\begin{subfigure}[b]{0.3\textwidth}
			\includegraphics[width=\textwidth,trim = 20mm 20mm 20mm 20mm, clip]{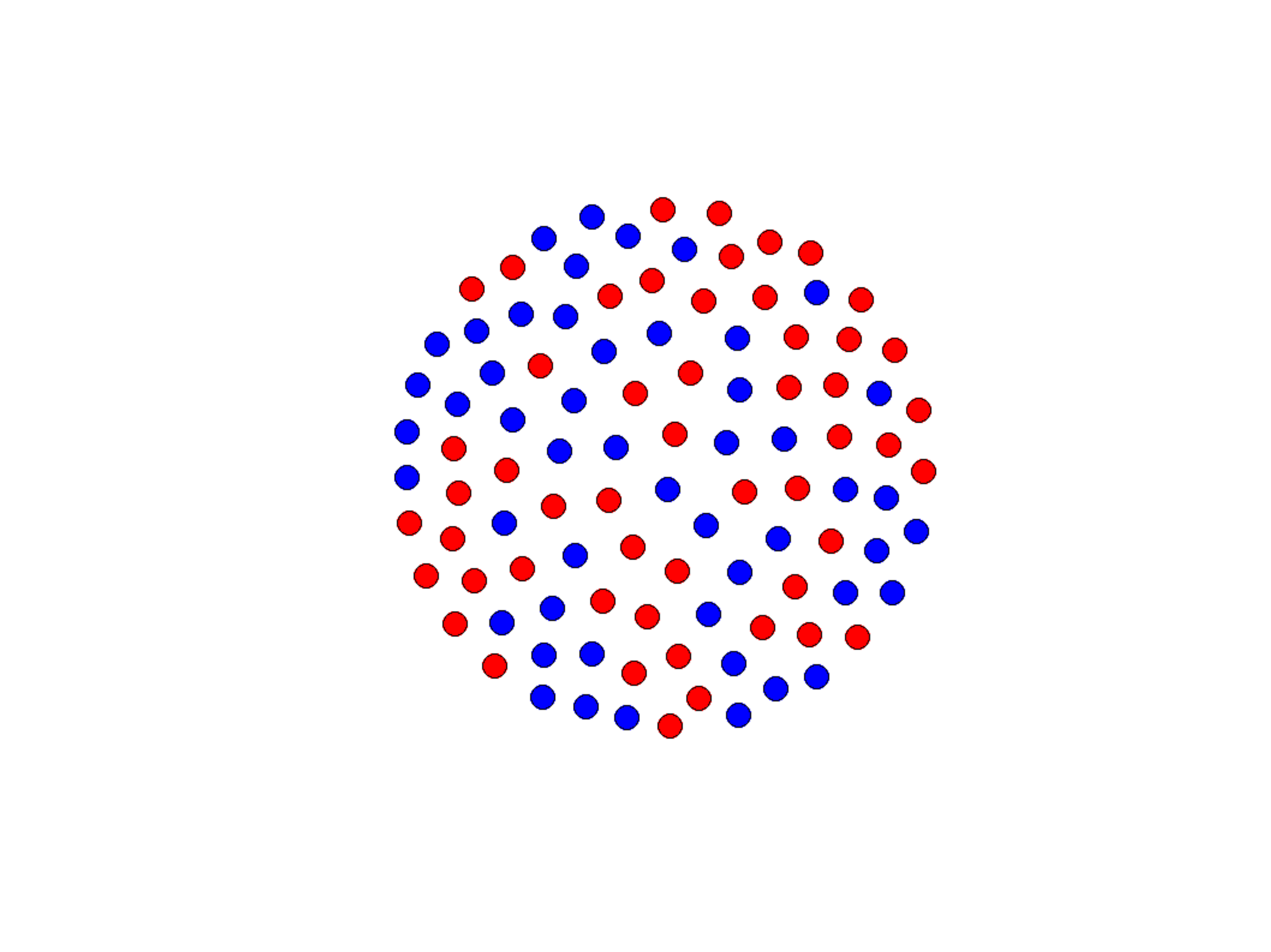}
			\caption{Co-occurrence}
			\label{F:Senate_O}
		\end{subfigure}%
		~ 
		\begin{subfigure}[b]{0.3\textwidth}
			\includegraphics[width=\textwidth,trim = 20mm 20mm 20mm 20mm, clip]{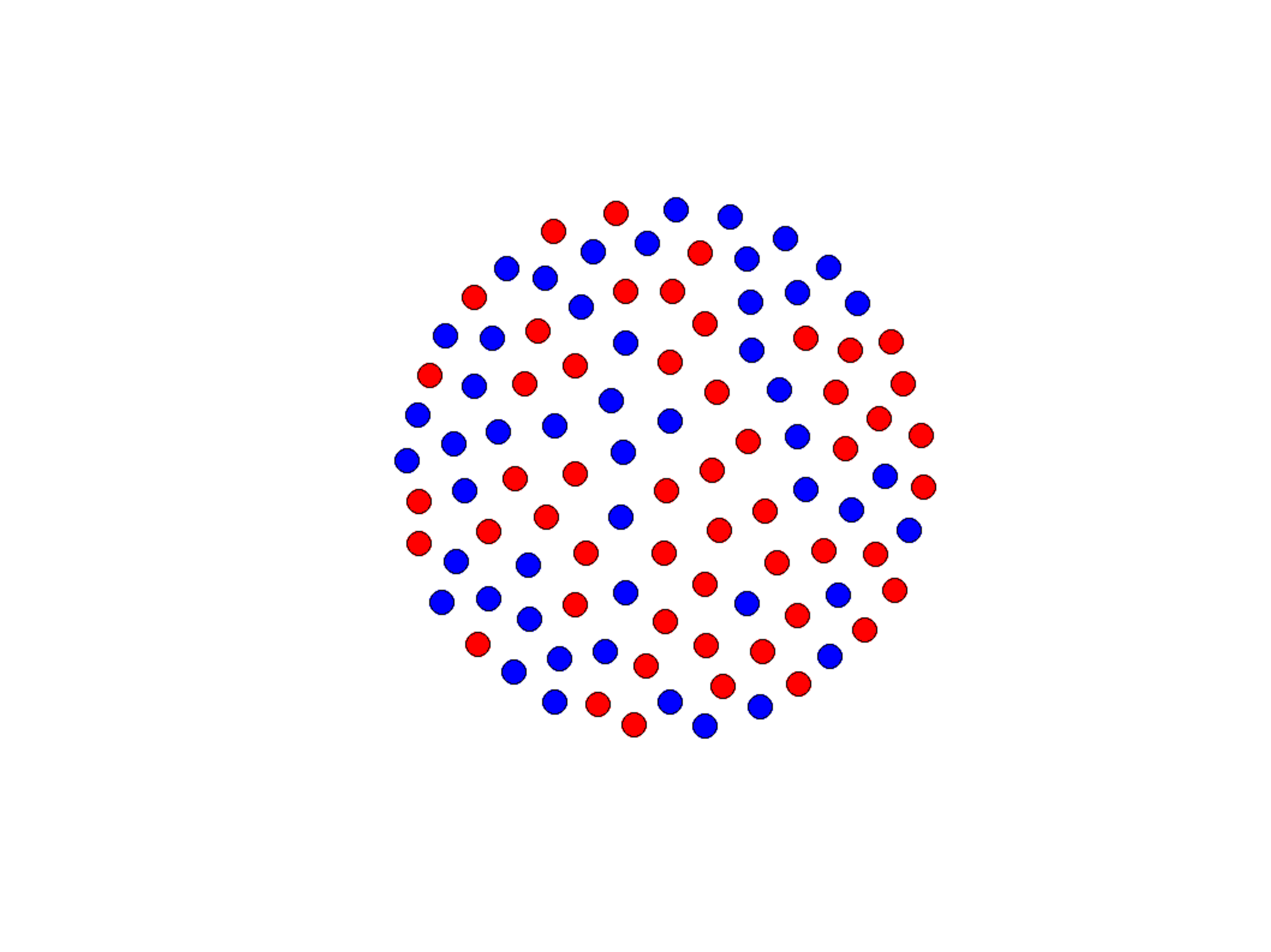}
			\caption{Half Weight Index }
			\label{F:Senate_H}
		\end{subfigure}
		~
		\begin{subfigure}[b]{0.3\textwidth}
			\includegraphics[width=\textwidth,trim = 20mm 20mm 20mm 20mm, clip]{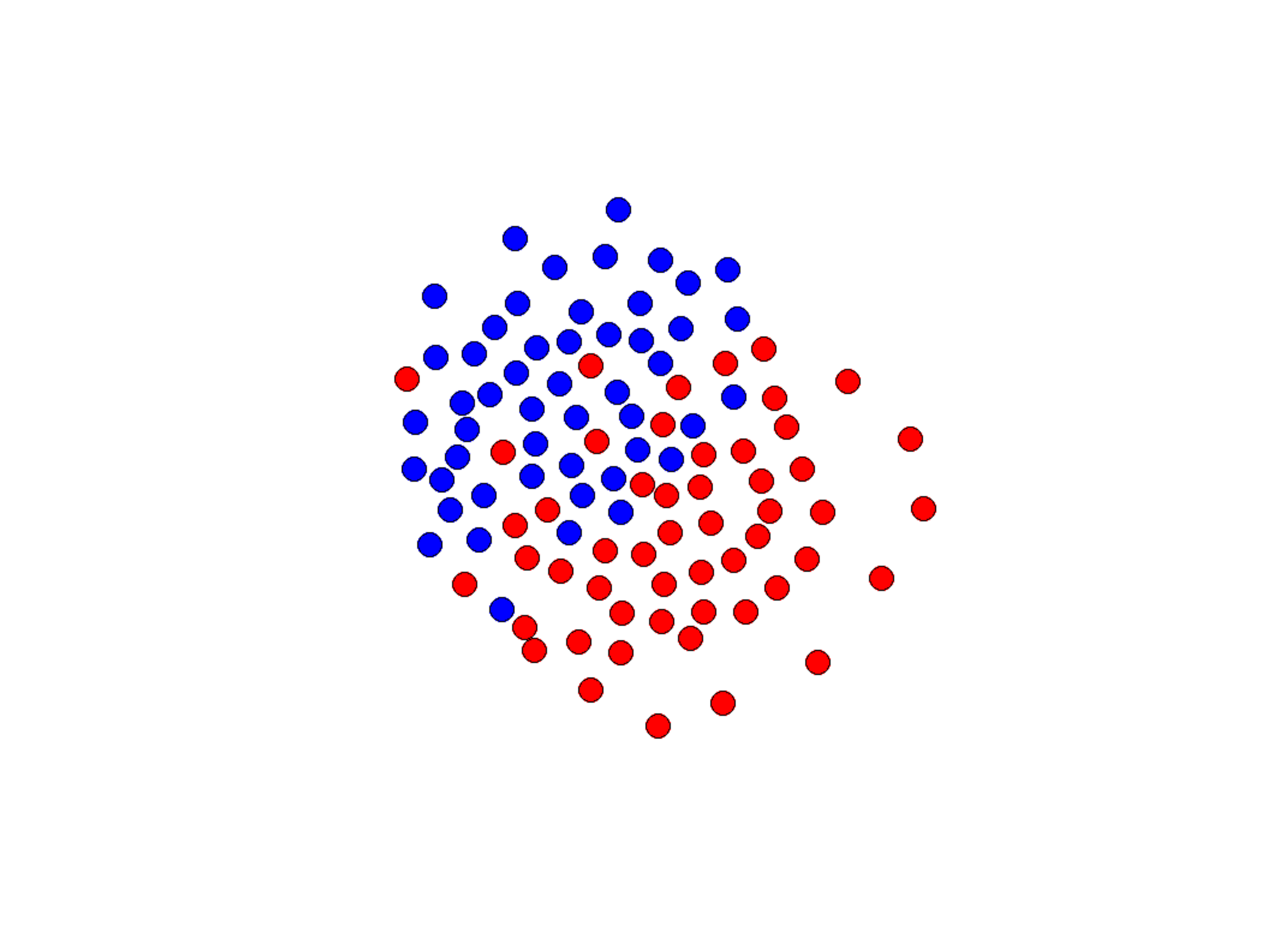}
			\caption{Hub Model }
			\label{F:Senate_A}
		\end{subfigure}
		\caption{Comparison of Estimation Techniques for the 110$^{th}$ Senate}\label{F:Senate}
	\end{figure}

	In Figure \ref{F:Senate}, we plot the co-occurrence matrix, half-weight index, and the adjacency matrix of the Hub Model using the force directed graph drawing technique of Fruchterman-Reingold.  Each Senator is represented as a node where the color of the node represents the Senator's official political party. Red nodes represent Republicans while blue nodes represent Democrats. We omit the links between nodes in the three sub-figures, since the networks constructed by the co-occurrence matrix and half-weight index are almost complete graphs. The estimate $\hat{A}$ of HM in Figure \ref{F:Senate_A} shows a clearly clustered pattern of Republicans and Democrats. By contrast, the figures of co-occurrence matrix and half-weight index show no pattern. 
	
	This implies that the HM provides more meaningful information about the community structure of this dataset than classical measures.  This is further supported by the normalized cut value corresponding to the Senator's official party membership \cite{Shi00} .
	\begin{equation}\label{E:Normalized Cut}
	\frac{\sum\limits_{i\in C_1,j\in C_2} A_{ij}}{\sum\limits_{i,j\in C_1} A_{ij}}+\frac{\sum\limits_{i\in C_1,j\in C_2} A_{ij}}{\sum\limits_{i,j\in C_2} A_{ij}}.
	\end{equation} 
	
	\eqref{E:Normalized Cut} gives the normalized cut value for two communities $C_1$ and $C_2$.  Lower normalized cut values indicate stronger community differentiation.  For the Senate data, Table \ref{T:Normalized Cut} presents the values for each inferred network. The normalized cut value for HM is lower than the co-occurrence matrix and half weight index, which strengthens the visual intuition that the estimates from HM provide better distinction between communities.
	{\linespread{1.0}
		\begin{table}[hbt!]
			\begin{center}
				\begin{tabular}{|c| c|}
					\hline
					& Normalized Cut Value\\
					\hline
					
					Co-Occurrence Matrix    &  0.837   \\
					\hline
					
					Half Weight Index &  0.823  \\
					\hline
					Hub Model & 0.757 \\
					
					\hline
				\end{tabular}
			\end{center}
			\caption{Normalized Cut Ratios of 110$^{th}$ Senate for Different Inference Techniques}
			\label{T:Normalized Cut}
		\end{table}
	}

	\subsection{Dream of the Red Chamber}
	
	As noted by Kolaczyk \cite{Kolaczyk09}, a significant challenge with parameter estimation in the context of implicit networks is that for a real world dataset there is usually no way to verify the extent to which the estimate matches reality.  That is, there is no so-called ``ground truth'' or ``golden standard'' to evaluate the performance of the estimated results against.  Therefore, to test the performance of Hub Models, it is useful to analyze data about which there is some qualitative knowledge of the relationships between nodes. 
	
	To this end, we construct a dataset of  characters from the 18th century Chinese novel \textit{Dream of the Red Chamber}, also known as \textit{The Story of the Stone}.  Since novels contain a qualitative social structure that is familiar to readers, the results of quantitative analysis can be compared to this standard.  This novel is used here because the relationships between the characters are subtle and complex. In this way, the story presents a challenge to the task of estimating the relationships between the characters. 
	
	Traditional approaches to building datasets from novels require carefully reading the text and identifying dyadic interactions between characters based on criteria established by the researchers, e.g., character $A$ has a conversation with character $B$ \cite{MacCarron13}. This method may construct high quality datasets; however, in order to identify the dyadic interaction, it requires readers who are familiar with the languages used in the novel and who have time to build the datasets.  Since \textit{Dream of the Red Chamber} is written in classical Chinese and the English translation runs over 2,600 pages, directly generating the dataset would be excessively time consuming.
	
	To overcome the size of the novel, we built the dataset by employing an automatic text mining technique. We define a group as characters who co-occur in the same paragraph. The novel can be text-mined automatically since character names and paragraph markers are easily identified.  Paragraphs with no characters named in them are ignored.
	
	We analyze the relationships of 29 important characters. The character names presented here are based on the original pinyin pronunciations and the David Hawkes translation \cite{Hawkes74}.  A Chinese version of the novel was used for text-mining. 
	
	This complete novel contains 120 chapters, but we focus on the first 80 chapters because it is commonly believed that the last 40 chapters are written by a different author and may not reflect the original themes of the novel.  The resulting dataset has 1,389 observations of groups containing at least one of the 29 characters.
	
\begin{figure}[H]
		\centering
		\begin{subfigure}[b]{0.3\textwidth}
			\includegraphics[width=\textwidth]{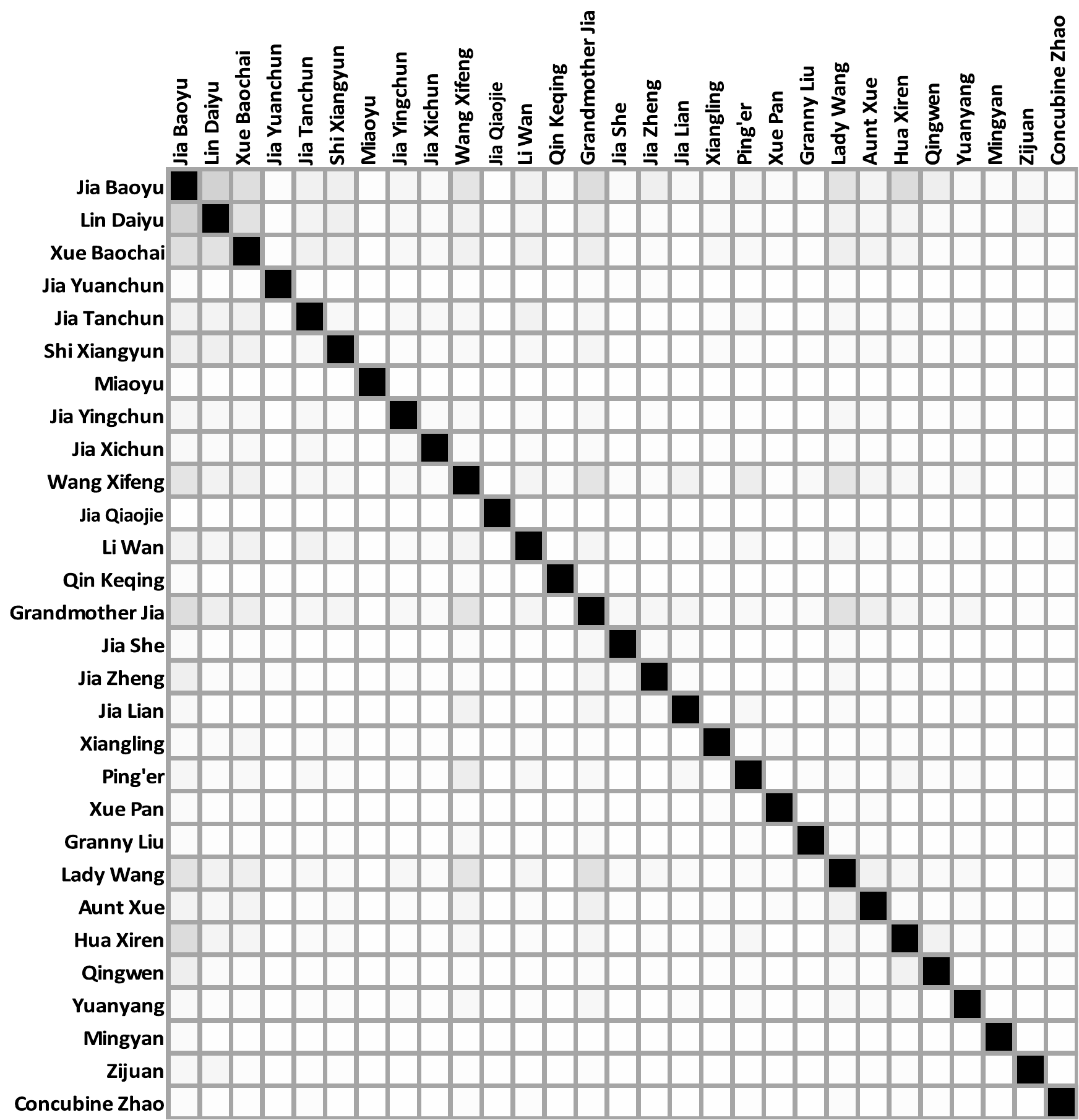}
			\caption{Co-Occurrence Matrix}
			\label{F:DRC_O_names}
		\end{subfigure}
		~
		\begin{subfigure}[b]{0.3\textwidth}
			\includegraphics[width=\textwidth]{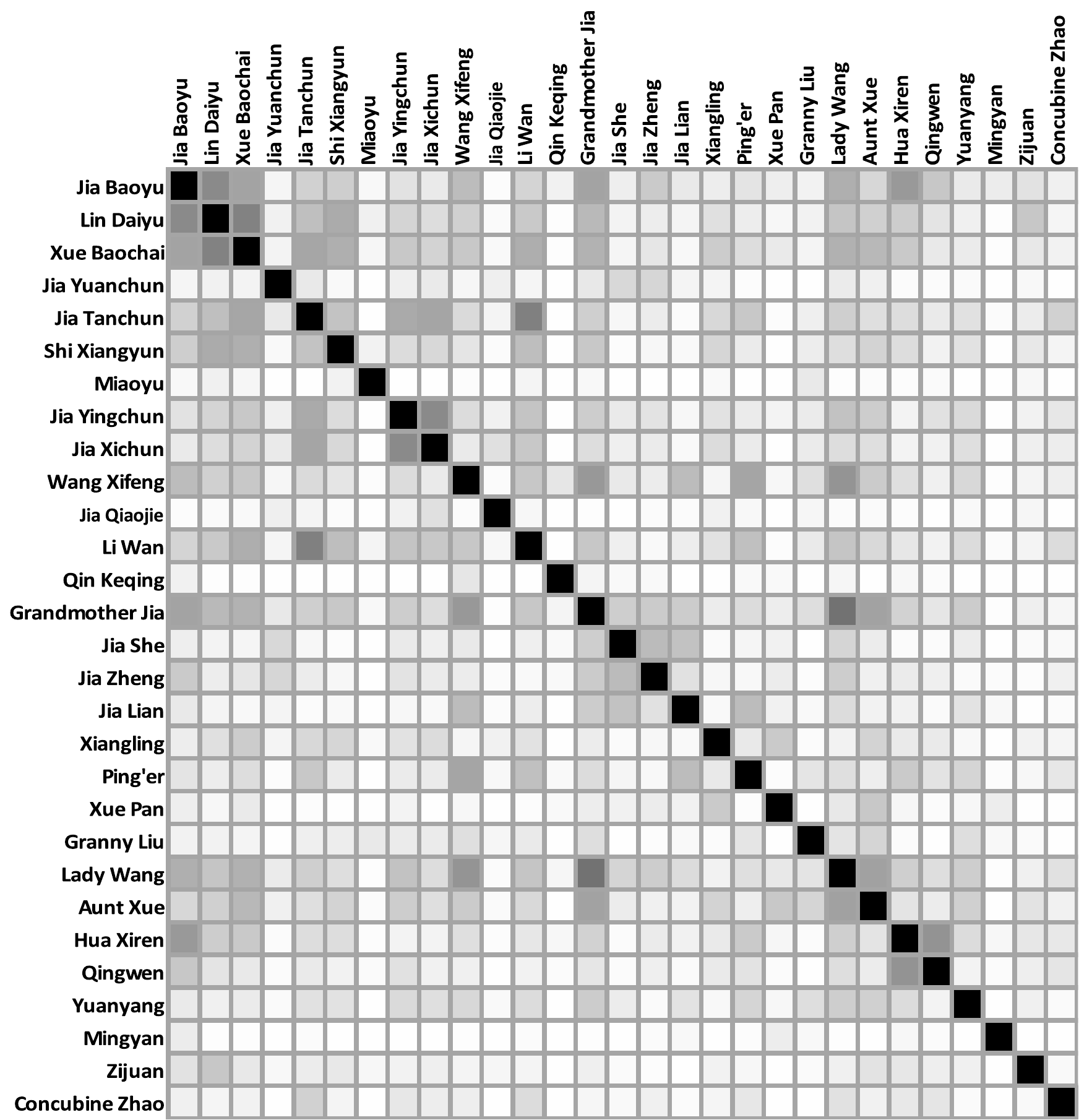}
			\caption{Half Weight Index}
			\label{F:DRC_H_names}
		\end{subfigure}
		\\
		\begin{subfigure}[b]{0.3\textwidth}
			\includegraphics[width=\textwidth]{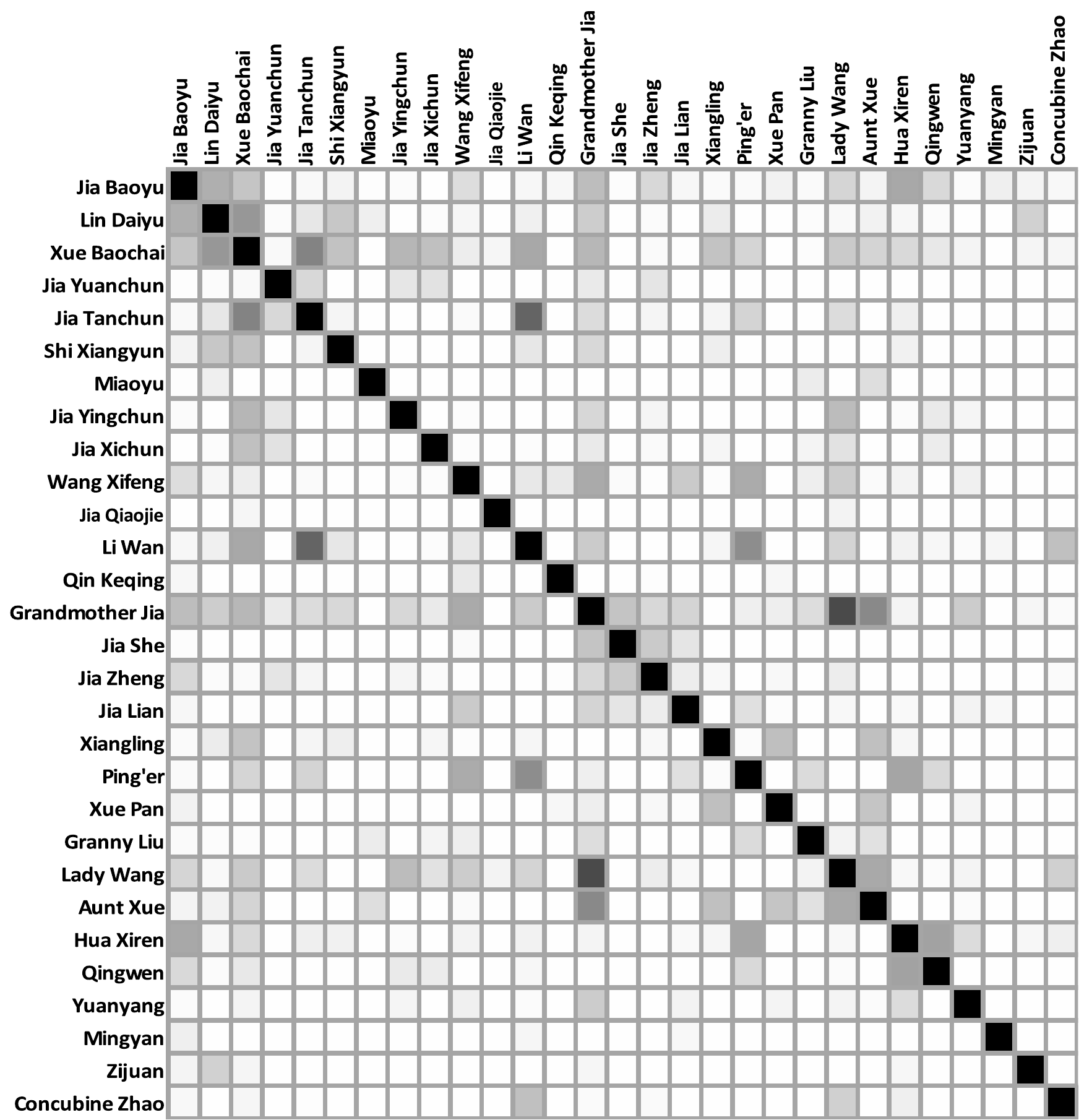}
			\caption{HM Adjacency Matrix}
			\label{F:DRC_A_names}
		\end{subfigure}
		\caption{Comparison of Estimation Techniques for \textit{Dream of the Red Chamber}}\label{F:DRC}
	\end{figure}

	Figure \ref{F:DRC} presents the results of the three techniques for estimating social structure discussed throughout this paper.  In this section, we employ an alternative visualization technique for the relationships between individual nodes.  The adjacency matrix, $A$ is represented as an $n \times n$ grid where the $i^{th} \times j^{th}$ cell represents the relationship $A_{ij}$.  The strength of a relationship is represented by the cell's color.  Nodes with weak relationships have light cells while nodes with strong relationship have dark cells.  Cells representing relationships of intermediate strength are shaded along the gray scale.
	
	This visualization demonstrates another difference in the performance of the techniques.  The co-occurrence matrix estimates all relationships as being very weak and it is difficult to differentiate strong relationships from the absence of a relationship.  The half-weight index presents a much stronger set of relationships but there is evidence of relationships which have been imputed transitively. In general, HM returns a much sparser network where the strength of relationships demonstrates higher contrast.  This tendency towards sparsity is discussed in more detail in Section \ref{S:Discussion}.

	The EM algorithm of HM provides very stable solutions. By selecting multiple starting points, we find that the adjacency matrix (Figure \ref{F:DRC_A_names}) is repeatedly returned as the most likely parameter of the observed data.
	
	The standard deviation of the parameters of the Hub Model was estimated using the bootstrap technique.  In general, the standard deviation was low. This was particularly true for $\hat{\rho}$ where the maximum standard deviation was 0.0173. Table \ref{T:DRC_Percentiles} presents the standard deviation of $\hat{A}$ at different percentiles.
	{\linespread{1.0}
		\begin{table}[hbt!]
			\begin{center}
				\resizebox{5in}{!}{
					\begin{tabular}{|c| c c c c c c c|}
						\hline
						Percentile & Max & 95 \% & 75 \% & Med & 25 \% & 5 \% & Min \\
						\hline
						StDev &         0.2696 & 0.1025 & 0.0374 & 0.0100 & 0.0000 & 0.0000 &  0.0000 \\
						\hline
					\end{tabular}
				}
			\end{center}
			\caption{Percentiles of Standard Deviation in $\hat{A}$ estimated by HM for \textit{Dream of the Red Chamber}}
			\label{T:DRC_Percentiles}
		\end{table}
	}
	
	
	One of the main themes of the \textit{Dream of the Red Chamber} is the love story surrounding the protagonist Jia Baoyu (1st character in Figure \ref{F:DRC_A_names}) and two potential fiancees.  These are the sickly Lin Daiyu (2nd character) and the ``ideal'' Xue Baochai (3rd character).  Although Jia Baoyu shares a special bond with Lin Daiyu and has no significant emotional connection to Xue Baochai, he is ultimately tricked into marrying Xue Baochai.  In Table \ref{T:DRC_Main}, we present the relationships between these two girls and the other characters as estimated by the co-occurrence matrix, half weight index, and HM.
	
	From the novel, Lin Daiyu is a sensitive and sickly girl who prefers to be alone. By contrast, Xue Baochai is a social and calculating girl. She is extremely good at interpersonal communication especially with the protagonist's mother (Lady Wang) and grandmother (Grandmother Jia). These significantly different personalities are clearly represented by the HM estimator while the other estimators do not identify this difference. 
	
	Xue Baochai generally has much stronger relationships with other characters, except for three: Jia Baoyu (the protagonist), Miaoyu (a nun with a very similar personality to Lin Daiyu) and Zijuan (a maid of Lin Daiyu). The co-occurrence matrix and half weight index fail to identify such a clear pattern. 
	{\linespread{1.0}
		\begin{table}[H]
			\begin{center}
				\resizebox{6in}{!}{
					\begin{tabular}{| c | c c | c c | c c |}
						\hline
						& \multicolumn{2}{c|}{Co-Occurrence Matrix ($O$)} & \multicolumn{2}{c|}{Half Weight Index ($H$)} & \multicolumn{2}{c|}{Hub ($\hat{A}$)}\\
						\hline
						& Lin   & Xue      & Lin   & Xue      & Lin & Xue  \\
						& Daiyu & Baochai  & Daiyu & Baochai  & Daiyu & Baochai\\
						\hline
						Jia Baoyu &   0.1728 &	0.1274 &	0.4563 &	0.3587 &	0.3113 &	0.2258 \\
						Lin Daiyu &   1.0000 &	0.1109 &  1.0000 &	0.4866 &	1.0000 &	0.4072 \\
						Xue Baochai & 0.1109 &	1.0000 &	0.4866 &	1.0000 &	0.4072 &	1.0000 \\
						Jia Yuanchun &0.0072 &	0.0050 &	0.0531 &	0.0449 &	0.0156 &	0.0228 \\
						Jia Tanchun  & 0.0439 &	0.0533 &	0.2490 &	0.3482 &	0.0915 &	0.4848 \\
						Shi Xiangyun & 0.0590 &	0.0490 &	0.3273 &	0.3119 &	0.2194 &	0.2365 \\
						Miaoyu       & 0.0072 &	0.0036 &	0.0552 &	0.0337 &	0.0597 &	0\\
						Jia Yingchun & 0.0252 &	0.0274 &	0.1667 &	0.2141 &	0 &	0.2846 \\
						Jia Xichun   & 0.0187 &	0.0202 &	0.1313 &	0.1692 &	0.0102 &	0.2461 \\
						Wang Xifeng  & 0.0497 &	0.0526 &	0.1840 &	0.2131 & 	0.0317 &	0.0697 \\
						Jia Qiaojie  & 0.0022 &	0.0022 &	0.0170 &	0.0208 &	0 &	0.0348 \\
						Li Wan       & 0.0367 &	0.0482 &	0.2086 &	0.3160 &	0.0580 &	0.3384 \\
						Qin Keqing   & 0.0007 &	0.0007 &	0.0052 &	0.0062 &	0 &	0 \\
						\hline
						Grandmother Jia    & 0.0655 &	0.0648 &	0.2725 &	0.2985 &	0.1925 &	0.2820 \\
						\hline
						Jia She      & 0.0065 &	0.0043 &	0.0449 &	0.0357 &	0 &	0 \\
						Jia Zheng    & 0.0122 &	0.0144 &	0.0701 &	0.0952 &	0.0143 &	0.0174 \\
						Jia Lian     & 0.0072 &	0.0036 &	0.0423 &	0.0245 &	0.0002 &	0.0073 \\
						Xiangling    & 0.0180 &	0.0252 &	0.1185 &	0.1961 &	0.0741 &	0.2344 \\
						Ping'er      & 0.0122 &	0.0209 &	0.0668 &	0.1306 &	0.0016 &	0.1643 \\
						Xue Pan      & 0.0043 &	0.0101 &	0.0292 &	0.0809 &	0 &	0  \\
						Granny Liu   & 0.0072 &	0.0050 &	0.0493 &	0.0411 &	0.0101 &	0.0113 \\
						\hline
						Lady Wang    & 0.0490 &	0.0590 &	0.2248 &	0.3037 &	0.0224 &	0.2065 \\
						\hline
						Aunt Xue     & 0.0302 &	0.0396 &	0.1806 &	0.2750 &	0.0479 &	0.1657 \\
						Hua Xiren    & 0.0403 &	0.0389 &	0.1938 &	0.2105 &	0.0283 &	0.1469 \\
						Qingwen      & 0.0166 &	0.0115 &	0.1020 &	0.0829 &	0.0155 &	0.0886 \\
						Yuanyang     & 0.0086 &	0.0101 &	0.0556 &	0.0763 &	0  &	0.0430 \\
						Mingyan      & 0.0007 &	0.0007 &	0.0053 &	0.0064 &	0  &	0  \\
						Zijuan       & 0.0317 &	0.0108 &	0.2184 &	0.0888 &	0.1775 &	0.0376 \\
						Concubine Zhao & 0.0050 &	0.0058 &	0.0361 &	0.0495 &	0 &	0.0338 \\
						\hline                 
					\end{tabular}
				}
			\end{center}
			\caption{Relationships of Lin Daiyu and Xue Baochai to other characters in \textit{Dream of the Red Chamber}}
			\label{T:DRC_Main}
		\end{table}
	}
	
	Figure \ref{F:DRC} and Table \ref{T:DRC_Main} show that even when the model assumption is not necessarily valid, important distinctions can be drawn from the Hub Model which are not possible with the other techniques.
	
	\subsection{North American Flora}\label{S:Flora}
	
	In the previous examples, we have worked with datasets which are essentially from the social sciences.  However, we believe that Hub Models are useful in other situations where observations are the result of nodes coalescing around a single node or observations are the result of some resource dispersing outward from a single node to multiple nodes.
	
	As a demonstration of how this kind of data can be used to estimate the relationship between different regions, we use a dataset from the University of California Irvine Machine Learning Repository which had been extracted and encoded from the USDA plants database \cite{Hamalainen08}.  34,781 plant species or geneses are included in the dataset.  For each plant, the dataset indicates which of 68 areas the plant is found in.  These areas include all United States states, Canadian provinces and territories, along with the Virgin Islands, Puerto Rico, Greenland, and St. Pierre and Miquelon (islands off the northeast coast of Canada).  For simplicity, we will refer to these areas as \textit{states}.
	
	We would expect that contiguous states would tend to have many flora in common while states which are far apart would be less likely to share common flora.  For example, Connecticut and Massachusetts are small states which share a common boarder; therefore, we would expect them to appear together in the regions of many plants.  Conversely, California and Greenland are very far apart and at different latitudes; therefore, we would expect a weak relationship.
	
	Of course, the map of North America is well known and our objective here is not to compare the Hub Model to spacial modeling.  Instead, we are using the regions of North America as a proxy for a system of distribution.  
	
	To demonstrate the ability of the Hub Model to capture the connections between states, we split them into 13 different regions.  The United States is identified by the 9 divisions of the US Census Bureau.  The Canadian provinces are identified by three regions.  The final region includes islands in the Atlantic ocean which are not included in any other region.
	
	\begin{figure}[H]
		\centering
		\fbox{
			\begin{subfigure}[b]{0.2\textwidth}
				\includegraphics[width=\textwidth,trim = 85mm 40mm 73mm 40mm, clip]{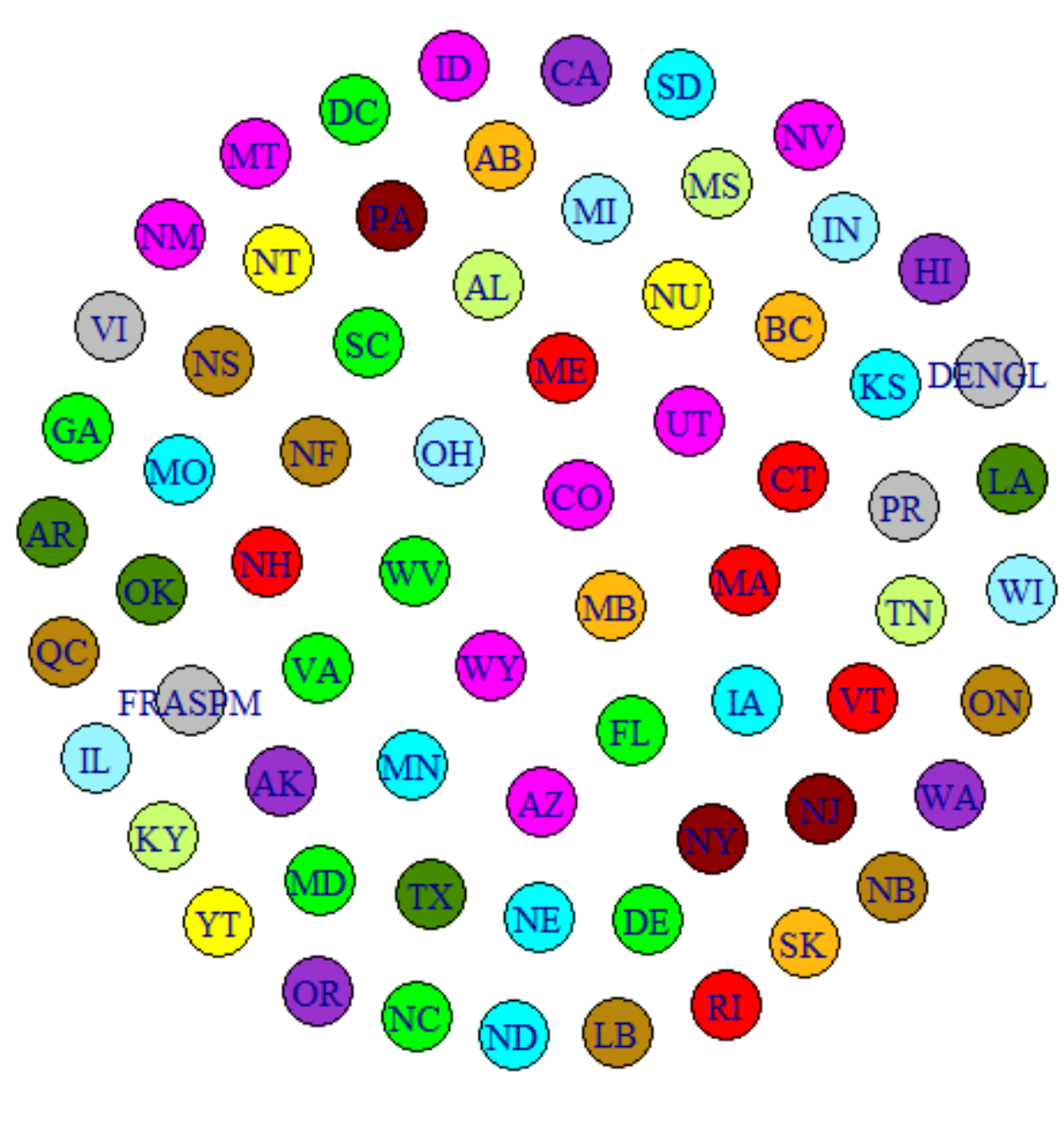}
				\caption{Co-Occurrence}
				\label{F:Plant_O}
			\end{subfigure}
		}
		~
		\fbox{
			\begin{subfigure}[b]{0.2\textwidth}
				\includegraphics[width=\textwidth,trim = 85mm 40mm 73mm 40mm, clip]{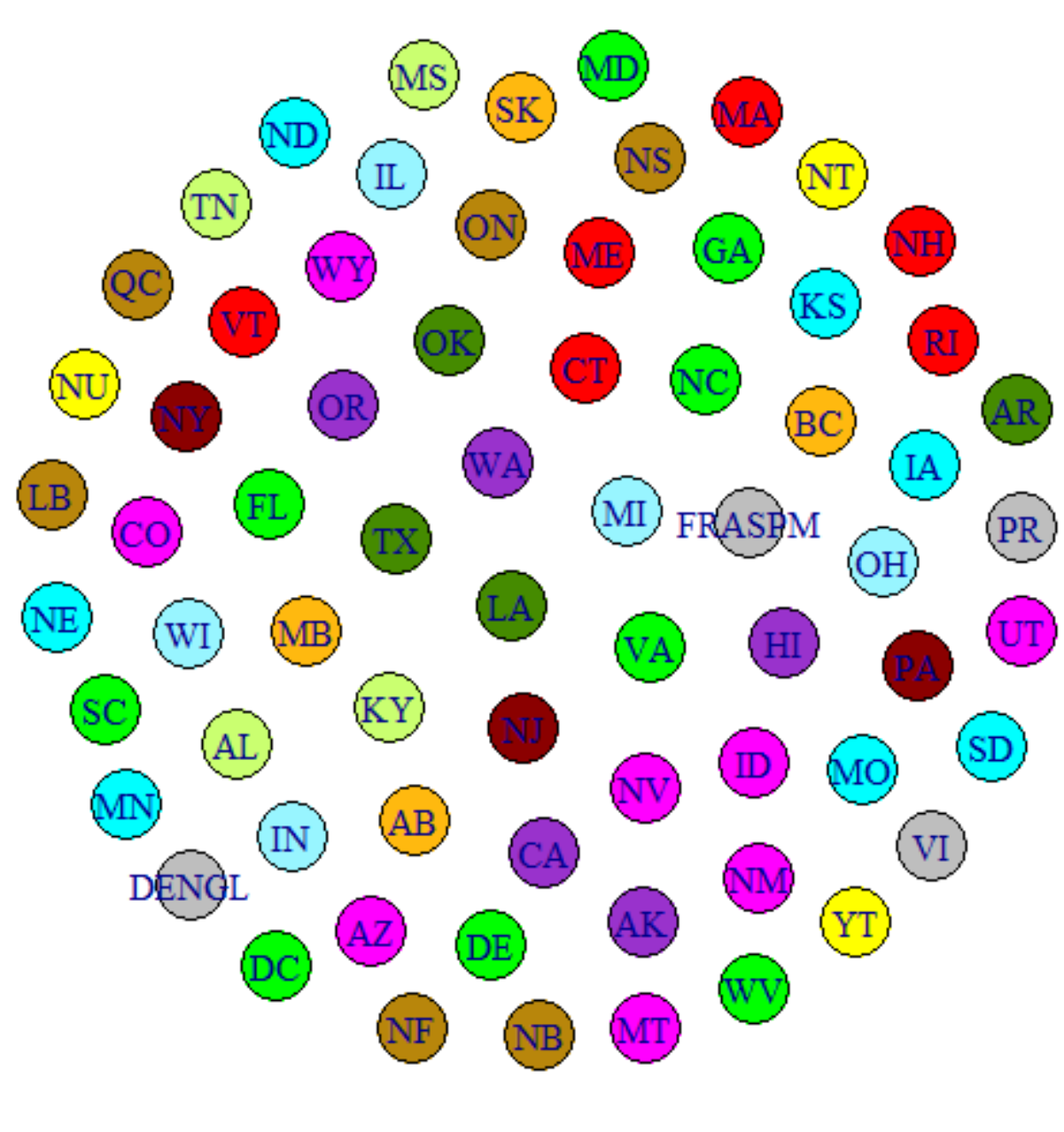}
				\caption{Half Weight Index}
				\label{F:Plant_H}
			\end{subfigure}
		}
		~
		\fbox{
			\begin{subfigure}[b]{0.2\textwidth}
				\includegraphics[width=\textwidth,trim = 86mm 40mm 73mm 40mm, clip]{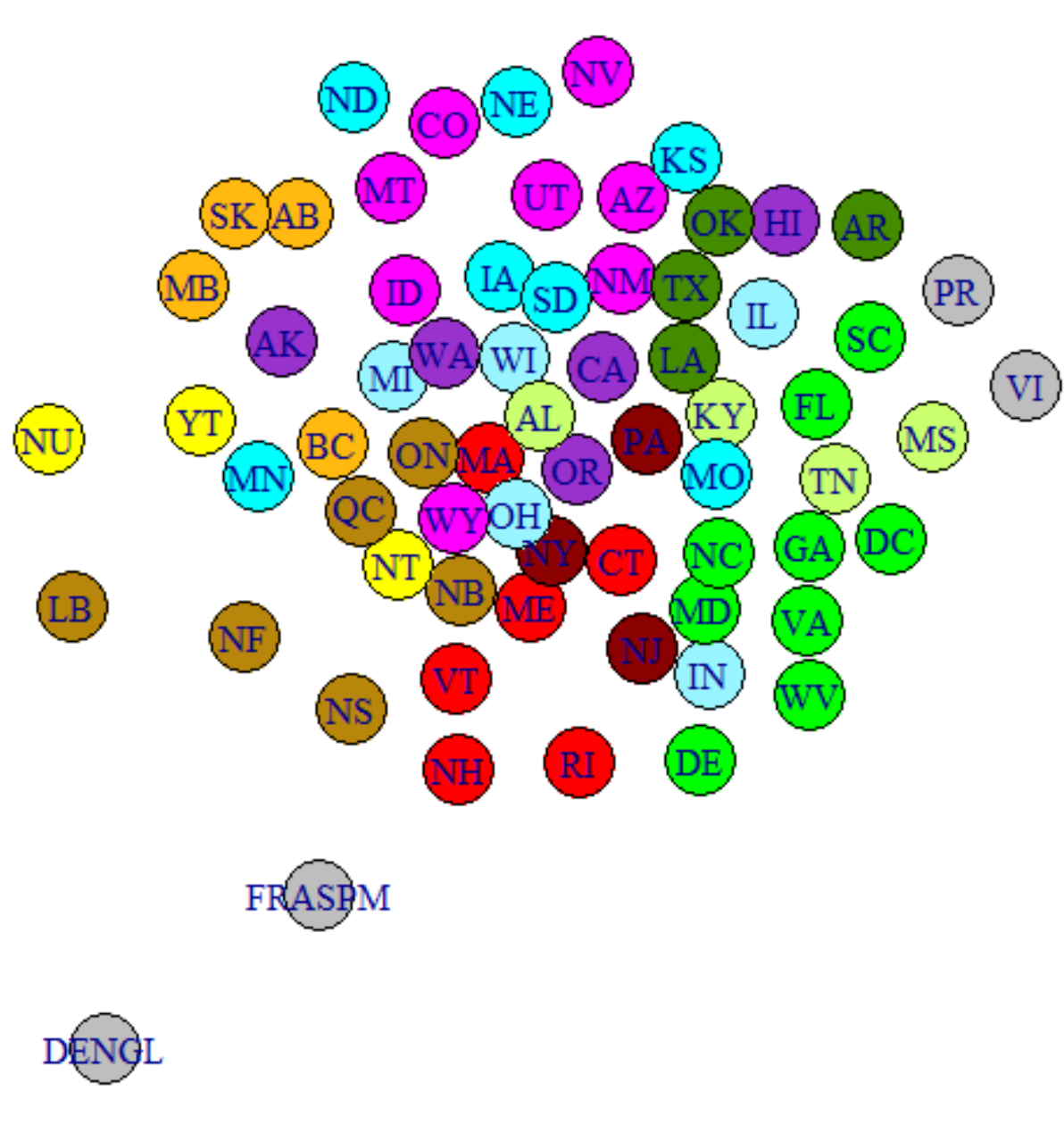}
				\caption{Hub Model}
				\label{F:Plant_A}
			\end{subfigure}
		}
		\\
		\begin{subfigure}[b]{.7\textwidth}
			\includegraphics[width=\textwidth]{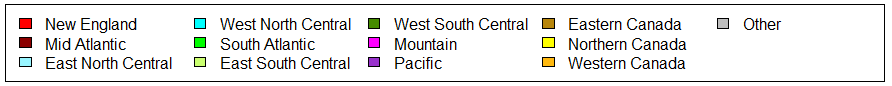}
			\label{F:Plant_ledgend}
		\end{subfigure}
		\caption{Adjacency Matrix Estimates for North American Flora Data}\label{F:Plant}
	\end{figure}
	
	The HM graph is striking in how closely states in the same region are grouped.  Southern states are generally on the right side of Figure \ref{F:Plant_A} while northern states are generally to the left.  Eastern states are generally at the bottom of the figure while western states are generally at the top.  Atlantic islands are on the outer edge of the plot.
	
	In Figures \ref{F:Plant_O} and \ref{F:Plant_H}, there is almost no distinction between the organization of the states.  This suggests that even in situations where the data does not clearly conform to the Hub Model assumption that valuable information about the relationship between nodes can be identified.
	
	\section{DISCUSSION}\label{S:Discussion}
	
	\subsection{Self-sparsity}
	
	In Section \ref{S:Simulation}, we introduced a property of the Hub Model estimators which we refer to as \textit{self-sparsity}.  When $T$ is small relative to $n$, the model tends to produce a sparse adjacency matrix. Rabbat et al.\cite{Rabbat08} observed similar behavior in their research.  This sparsity in $A$ is achieved without any penalty in the log-likelihood, hence the name.  
	
	To begin, observe that the true probability of co-occurrence is related to $\{A,\rho\}$ by the following equation: 
	\begin{equation}\label{E:Sparsity}
	\mathbb{P}(v_i \textnormal{ and } v_j \textnormal{ co-occur})=\sum_{k=1}^n \rho_k A_{ki} A_{kj}.
	\end{equation}
	
	Suppose that there is a pair of nodes, $\{v_i,v_j\}$, for which the probability of co-occurrence is exactly zero. Equation \eqref{E:Sparsity} implies that:
	\begin{align*}
	\rho_k A_{ki} A_{kj} &= 0 \quad \forall k.
	\end{align*}
	
	Hence, for every $k$, at least one of the following is true: $\rho_k=0$, $A_{ki}=0$, or $A_{kj}=0$.  At a minimum, this requires that there be $n$ elements of the parameters $\{A,\rho\}$ which are exactly equal to zero for every pair of nodes which fails to co-occur.  
	
	Clearly, $O_{ij}=0$ implies $A_{ij}=0$.  However, self-sparsity shows that the absence of co-occurrence contains even more information than just the relationship between two nodes.  Absence of co-occurrence means that no member of the population chooses to simultaneously interact with both nodes.
	
	However, the question remains as to why self-sparsity occurs in the estimation of $\{A,\rho\}$ when $T$ is small relative to $n$.  We observe that sparsity in $\hat{A}$ or $\hat{\rho}$ is a consequence of the EM-algorithm of HM.  
	
	First, it is easy to check that zero is an absorbing state for $\hat{A}$ and $\hat{\rho}$ in the EM-algorithm.  If at the $m^{th}$ iteration, $\hat{\rho}_i^{(m)}=0$, then by  \eqref{E:ProbKisCenter}  $\mathbb{P}(S_i=1|G^{(t)})=0$ for all $i$, and by \eqref{E:HM rho_hat} $\hat{\rho}_i^{(m+1)}=0$.  Therefore, $\hat{\rho}_i=0$ is an absorbing state.  For a similar reason, $\hat{A}_{ij}=0$ is also an absorbing state.
	
	Note that in the EM-algorithm we set $\hat{A}_{ij}^{(m)}=0$ when it is below a certain threshold.  But why the estimate approaches zero is not fully understood.  This aspect of the model will be explored in future work.

	\subsection{Identifiability}
	
	Recall from Section \ref{S:Ident} that when we allow $A$ to be asymmetric, the model is not identifiable.  That is, the following condition is not satisfied:
	\begin{equation}
	\mathbb{P}(G=g|A,\rho)=\mathbb{P}(G=g|A^*,\rho^*) \hspace{2mm} \forall g \implies \{A,\rho\}=\{A^*,\rho^*\}.
	\end{equation} 
	
	In Section \ref{S:Ident}, we showed that symmetry ($A_{ij}=A_{ji}$) is an identifiability condition.  In this subsection we explore identifiability in more detail.
	
\begin{figure}[H]
		\centering
		\includegraphics[width=0.4\textwidth]{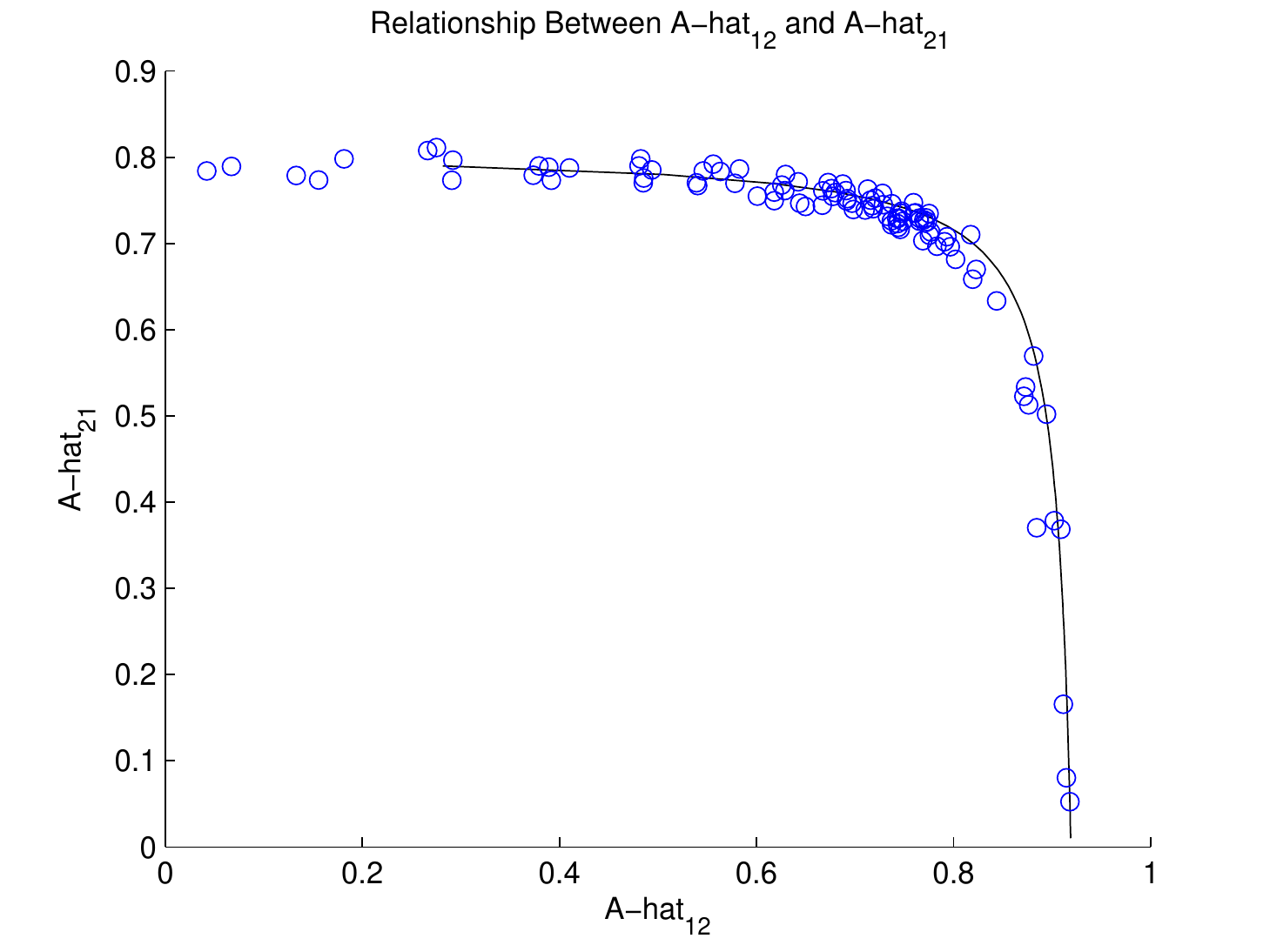}
		\caption{Nonlinear Relationship of Symmetric Elements for $\hat{A}$}
		\label{F:A12}
	\end{figure}

	We give a simple counterexample in Section \ref{S:Ident} to demonstrate that without any constraints on $A$ and $\rho$, the model is not identifiable. Here we will randomly select parameters to explore the issue in general. We randomly generate an asymmetric adjacency matrix $\{A,\rho\}$ with $n=4$ (see Table \ref{T:A12}).  Note that for the asymmetric case, $n=4$ is the smallest population size where the  number of possible groups exceeds the number of parameters to estimate.  The number of observed groups is set high ($T=100,000$) to ensure good performance of the algorithm.  We ran Algorithm 1 100 times, and obtained 100 different estimators with the same or very close likelihoods. Figure \ref{F:A12} gives a scatterplot of the one hundred pairs of $\{\hat{A}_{1,2},\hat{A}_{2,1}\}$ indicated by blue circles.  This plot clearly demonstrates a non-linear relationship between these two values where increases in one are associated with decreases in the other.
	{\linespread{1.0}
		\begin{table}[H]
			\begin{center}
				\begin{tabular}{| l | c | l l l l |}
					\cline{2-6}
					\multicolumn{1}{c}{}  &  \multicolumn{5}{|c|}{$A_{ij}$}\\
					\cline{3-6}
					\multicolumn{1}{c}{}  & \multicolumn{1}{|c}{}    & \multicolumn{4}{|c|}{j}\\
					\hline
					$\rho_i$ &  i & 1 & 2 & 3 & 4 \\
					\hline
					0.5499 & 1 & 1.0000 & 0.7854 & 0.9063 &  0.7957 \\
					0.3269 & 2 & 0.7032 & 1.0000 & 0.8324 &  0.5885 \\
					0.1016 & 3 & 0.9464 & 0.8817 & 1.0000 &  0.9334 \\
					0.0216 & 4 & 0.7452 & 0.8594 & 0.9478 &  1.0000 \\
					\hline                 
				\end{tabular}
			\end{center}
			\caption{True Adjacency Matrix in Identifiability Example}
			\label{T:A12}
		\end{table}
	}
	
	We now derive the theoretical relationship between symmetric elements of the adjacency matrix. As in Section \ref{S:Ident}, let $g^x$ and $g^y$ represent the observed groups which contain only node $v_x$ and $v_y$ respectively.  Further, let $g^{xy}$ represent the group that is observed to contain only the pair $v_x$ and $v_y$.
	
	Under the Hub Models $\mathbb{P}(G=g^x)$ and $\mathbb{P}(G=g^x)$ are given by \eqref{E:Prob_gx} and \eqref{E:Prob_gy}.  In the asymmetric case, the probability of the pair is:
	\begin{equation}\label{E:gxy}
	\mathbb{P}(G=g^{xy})=\rho_x A_{xy} \prod_{j\ne \{x,y\}}(1-A_{xj}) + \rho_y A_{yx} \prod_{j\ne \{x,y\}}(1-A_{yj}).
	\end{equation}
	
	By simply reordering the terms of Equations \ref{E:Prob_gx} and \ref{E:Prob_gy}, they can be plugged back into Equation \ref{E:gxy} to give:
	\begin{equation}\label{E:gxy2}
	\mathbb{P}(G=g^{xy})=\frac{A_{xy}}{(1-A_{xy})} \mathbb{P}(G=g^x)  + 
	\frac{A_{yx}}{(1-A_{yx})} \mathbb{P}(G=g^y)
	\end{equation}
	
	By some simple algebra, we find the following relationship between the elements of the adjacency matrix:
	\begin{equation}\label{E:SymRule}	A_{xy}=\frac{\mathbb{P}(G=g^{xy})-A_{yx}\big[\mathbb{P}(G=g^{y})+\mathbb{P}(G=g^{xy})\big]}{\big[\mathbb{P}(G=g^{x})+\mathbb{P}(G=g^{xy})\big]-A_{yx}\big[\mathbb{P}(G=g^{x})+\mathbb{P}(G=g^{y})+\mathbb{P}(G=g^{xy})\big]}.
	\end{equation}
	
	Using \eqref{E:SymRule}, we can calculate the relationship between $A_{1,2}$ and $A_{2,1}$ from the example above.  This is represented in Figure \ref{F:A12} by the solid line.  Clearly the observed solutions are falling along this theoretical curve. 
	
	\section{CONCLUSION}
	
	To the best of our knowledge, Hub Models introduce an innovative approach to the task of implicit network inference.  By defining a model-based generating mechanism to link the latent network to observed grouped data and applying an EM algorithm, we are able to estimate the network using this model.
	
	Not only are the estimators easy to calculate in a reasonable amount of time, but the estimators have a practical interpretation.  The parameter $\rho$ measures the probability that a member of a population will form a group.  $A_{ij}$ measures the probability that a member of the population will be included in a group formed by node $v_i$.
	
	The Hub Models compare favorably against existing techniques.  Since the co-occurrence matrix and half weight index lack a generating mechanism to connect them to the observed grouped data, these measures often cannot detect important features of a network.
	
	By applying the Hub Model to the 110$^{th}$ United States Congress dataset, the 18$^{th}$ century Chinese novel \textit{Dream of the Red Chamber}, and a dataset of flora in North America, we demonstrate that the HM is able to detect important features in the relationships between nodes in complex situations.  We conclude by providing some initial insights into self-sparsity and the identifiability of HM. To fully understand these properties is an intriguing research topic and will be explored in our future works.  
	
	\par
	\vskip 14pt
	\noindent {\large\bf Acknowledgements}
	
	This work is partially supported by NSF DMS 1513004.
	\par
	


\begin{thebibliography}{99}
		
		\bibitem{Anandkumar15} Anandkumar, A., Foster, D.~P., Hsu, D., Kakade, S.~M., and Liu, Y. \emph{A spectral algorithm from latent dirichlet allocation}, Algorithmica, 72(1):193--214 (2015).
			
		\bibitem{Bejder98} Bejder, L., Fletcher, D., and Brager, S. \emph{A method for testing association patterns of social animals}, Animal Behavior, 56:719--725 (1998).
				
		\bibitem{Brent11} Brent, L. J.~N., Lehmann, J., and Ramos-Fernandez, G. \emph{Social network analysis in the study of nonhuman primates: A
			historical perspective}, American Journal of Primatology, 73:720--730 (2011).
				
		\bibitem{Cairns87} Cairns, S.~J. and Schwager, S.~J. \emph{A comparison of association indices}, Animal Behavior, 35 (1987).
				
		\bibitem{Carreira00} Carreira-Perpinan, M.~A. and Renals, S. \emph{Practical identifiability of finite mixtures of multivariate
			bernoulli distributions}, Neural Computation, 12:141--152 (2000).
						
		\bibitem{Choudhury10} M. Choudhury, W. A. Mason, J. M. Hofman, D. J. Watts \emph{Inferring Relevant Social Networks from Interpersonal Communication}, International World Wide Web Conference Committee, April 26-30, 2010.
		
		\bibitem{Colace15} Colace, F., De~Santo, M., Greco, L., Moscato, V., and Picariello, A. \emph{A collaborative user-centered framework for recommending items in online social networks}, Computers in Human Behavior (2015).
				
		\bibitem{Dice45} Dice, L.~R. \emph{Measures of the amount of ecological association between species}, Ecology, 26:297--302(1945).
				
		\bibitem{Fowler06a}	Fowler, J.~H. \emph{Connecting the congress: A study of cosponsorship networks},Political Analysis, 14(4):456--487 (2006a).
				
		\bibitem{Fowler06b}	Fowler, J.~H. \emph{Legislative cosponsorship networks in the u.s. house and senate},Social Networks, 28(4):454--465 (2006b).
			
		\bibitem{Freeman89}	Freeman, L.~C., White, D.~R., and Romney, A.~K. \emph{Research Methods in Social Network Analysis}, George Mason University Press (1989).
			
		\bibitem{Hamalainen08} Hamalainen, W. and Nykanen, M. \emph{Efficient discovery of statistically significant association rules}, Proceedings of the 8th IEEE International Conference on Data Mining, pages 203--212 (2008).
	
		\bibitem{Han11} Han, J., Kamber, M., and Pei, J. \emph{Data Mining: Concepts and Techniques}, Morgan Kaufmann (2011).
				
		\bibitem{Hawkes74} Hawkes, D. \emph{The Story of the Stone, or The Dream of the Red Chamber, Vol. 1:	The Golden Days}, Penguin Classics (1974).
				
		\bibitem{Hiller01} Hiller, F.~S. and Lieberman, G.~L. \emph{Introduction to Operations Research},McGraw-Hill (2001).
			
		\bibitem{Kolaczyk09} Kolaczyk, E.~D. \emph{Statistical Analysis of Network Data: Methods and Models},Springer (2009).
			
		\bibitem{MacCarron13} MacCarron, P. and Kenna, R. \emph{Viking sagas: Six degrees of icelandic separation-social networks	from the viking era},Significance, pages 12--17 (2013).
				
		\bibitem{McLachlan08} McLachlan, G.~J. and Krishnan, T. \emph{The EM Algorithm and Extensions},John Wiley and Sons, Inc (2008).
				
		\bibitem{Newman11} M. Newman \emph{Networks: An Introduction}, Oxford University Press, 2011.
				
		\bibitem{Rabbat08} M. Rabbat, M. Figueiredo, and R. Nowak \emph{Network inference from co-occurrences}, IEEE Transactions on Information Technology 54(9): 4053--4068 (2006)
		
		\bibitem{Shi00}	Shi, J. and Malik, J. \emph{Normalized cuts and image segmentation}, IEEE Trans. Pattern Anal. Mach. Intell., 22(8):888--905(2000).
		
		\bibitem{Teicher61}	Teicher, H. \emph{Identifiability of mixtures}, The Annals of Mathematical Statistics, 32(1):244--248 (1961).
		
		\bibitem{Voelkl11} Voelkl, B., Kasper, C., and Schwab, C. \emph{Network measures for dyadic interactions: Stability and reliability}, American Journal of Primatology, 73:731--740 (2011).
		
		\bibitem{Vretos12} Vretos, N., Nikolaidis, N., and Pitas, I. \emph{Video fingerprinting using latent dirichlet allocation and facial images}, Pattern Recognition, 45(7):2489--2498(2012).
		
		\bibitem{Wasserman94} Wasserman, S. and Faust, C. Social Network Analysis: Methods and Applications, Cambridge University Press (1994).
		
		\bibitem{Zachary77} Zachary, W. \emph{An Information Flow Model for Conflict and Fission in Small Groups}, Journal of Anthropological Research, Vol 33 (1977) pages 452-473.  
		
	\end{thebibliography}
\end{document}